%% file: issigi.tex
\RequirePackage{fix-cm}
\documentclass[natbib,smallextended]{svjour3}       
\smartqed  
\usepackage{graphicx}
\usepackage{aps-bibstyle}  
\usepackage[extd]{journalnames}
\usepackage{amsmath}
\usepackage{amssymb}
\usepackage{lscape}
\usepackage[bookmarks=true, bookmarksopen=true]{hyperref}
\journalname{Space Science Reviews}

\begin{document}

\title{Insights into planet formation from debris disks
}
\subtitle{II. Giant impacts in extrasolar planetary systems}
\titlerunning{Giant impacts in extrasolar planetary systems}        
\author{Mark C. Wyatt \and Alan P. Jackson}
\institute{M. C. Wyatt \at
              Institute of Astronomy, University of Cambridge,
              Madingley Road, Cambridge CB3 0HA, UK \\
              Tel.: +44 1223 337 517, Fax: +44 1223 337 523 \\
              \email{wyatt@ast.cam.ac.uk}            \\
           \and
           A. P. Jackson \at
              School of Earth and Space Exploration, Arizona State University,
              Tempe, AZ 85287, USA \\}

\date{Received: date / Accepted: date}
\maketitle

\begin{abstract}
{\it Giant impacts} refer to collisions between two objects each of which
is massive enough to be considered at least a planetary embryo.
The putative collision suffered by the proto-Earth that created the Moon is
a prime example, though most Solar System bodies bear signatures of such collisions.
Current planet formation models predict that an epoch of giant impacts may be inevitable,
and observations of debris around other stars are providing mounting evidence that
giant impacts feature in the evolution of many planetary systems.
This chapter reviews giant impacts, focussing on what we can learn about planet
formation by studying debris around other stars.
Giant impact debris evolves through mutual collisions and dynamical interactions with
planets.
General aspects of this evolution are outlined, noting the importance of the
collision-point geometry.
The detectability of the debris is discussed using the example of the Moon-forming
impact.
Such debris could be detectable around another star up to 10\,Myr post-impact,
but model uncertainties could reduce detectability to a few 100\,yr window.
Nevertheless the 3\% of young stars with debris at levels expected during terrestrial
planet formation provide valuable constraints on formation models;
implications for super-Earth formation are also discussed.
Variability recently observed in some bright disks promises to illuminate the evolution
during the earliest phases when vapour condensates may be optically thick and
acutely affected by the collision-point geometry.
The outer reaches of planetary systems may also exhibit signatures of giant impacts, 
such as the clumpy debris structures seen around some stars.
\end{abstract}

\input{issigi_intro}
\input{issigi_giss}
\input{issigi_observables}
\input{issigi_gideb}

\input{issigi_moon}

\input{issigi_comparison}
\input{issigi_larger}

\input{issigi_superearth}
\input{issigi_conclusions}

\begin{acknowledgements}
MCW is grateful for support from the European Union through ERC grant number 279973.
\end{acknowledgements}

\bibliographystyle{aps-nameyear}      
\bibliography{refs}                

\end{document}

%% file: issigi_intro.tex
\section{Introduction}
\label{s:giintro}

The majority of the processes that contribute to the formation of a planetary system
are thought to take place in the massive circumstellar disks that surround young stars
called protoplanetary disks.
Such disks last for up to around 10\,Myr before dispersing through mechanisms that are still
debated, leaving a planetary system and a debris disk \citep{Wyatt2015}.
The debris disk is made up of the components of the planetary system that are not massive enough
to be classed as planets, i.e., planetesimals, dust and gas \citep[see][]{Matthews2014}.
For example, the Asteroid Belt and Kuiper Belt account for the majority of Solar System's 
debris disk, along with the Zodiacal Cloud and the Oort Cloud
\citep[see e.g.,][]{Matthews2016}.

The motivation for saying that planet formation processes are largely complete by the time the
protoplanetary disk is dispersed is based on the absence of large quantities of gas and
dust at later stages from which to form planets \citep[e.g.,][]{Haisch2001}.
However, while the disappearance of the protoplanetary gas disk does make the formation of gas
giant planets like Jupiter rather problematic, this does not necessarily prevent the continued
growth of solid rocky or icy planets.
From an observational perspective we can only say that the mass present in less than cm-sized dust
cannot be significant after the protoplanetary disk has dispersed \citep{Panic2013}.
The majority of the solid mass in the system could instead be locked up in planets
(or planetary embryos) that continue to grow.
Indeed this is the basis of some models for planetary system formation and evolution
that aim to explain some debris disk observations \citep{Kenyon2008}.

Isotopic compositions indicate that the parent bodies of many of the meteorites formed within
the first few Myr of the Solar System's history \citep[e.g.][]{Kleine2009}, i.e., while the
protoplanetary disk would have still been present.
However, a lengthier timescale for the formation of solid rocky planets is inferred from isotopic
measurements of the terrestrial planets, which shows that they formed over a range of timescales,
with Earth taking up to $\sim 100$\,Myr to reach its final mass \citep[e.g.][]{Halliday2008}.
The long formation timescale for the Earth is readily understood within the context of favoured
models for planet formation.
In such models, the protoplanetary disk is suffused with planetesimals that grow via mutual
collisions to eventually become planets, with a system of terrestrial planets that bears many
similarities to our own inner Solar System a likely outcome for the right initial conditions
\citep[e.g.,][]{Hansen2009,Walsh2011}.
Of relevance to the current discussion is that planetesimal growth results in many planetary
embryos \citep{Kokubo1998}, and in the final stage the number of embryos is whittled down to a few as they
undergo repeated chaotic close encounters and eventually giant collisions, until the few
that remain are sufficiently separated to be dynamically stable \citep[e.g.,][]{Goldreich2004,Kenyon2006}.
One consequence of this scenario is that the formation of the Earth's moon in a giant impact
$\sim 100$\,Myr after the Solar System formed is not an atypical occurrence.

The detailed structures of other inner Solar System bodies also bear witness to a past epoch of
giant (i.e., planetary scale) impacts in the inner Solar System \citep[e.g.,][]{Benz1988,Marinova2011}.
There are more uncertainties about the exact formation mechanism of the outer Solar System
\citep{Goldreich2004,Helled2014}, but there are several lines of evidence to suggest that giant
impacts were also a feature of the more distant reaches of the Solar System
\citep[e.g.,][]{Safronov1972,Canup2005}.
There is also growing evidence for giant impacts having occurred recently in some extrasolar
planetary systems, from observations of what could be impact-generated debris
\citep{Song2005,Rhee2008}.
Stars with anomalously high levels of dust in their regions that would be analogous to
the location of terrestrial planets in the inner Solar System are often interpreted within
the context of terrestrial planet formation \citep[e.g.,][]{Lisse2008,Melis2010},
particularly if they are at an age at which planet formation models predict such
giant impacts to be taking place (i.e., $\ll 100$\,Myr).
Clearly, studying this debris and how frequently it is detected around nearby stars has
the potential to inform about terrestrial planet formation processes more generally, as
well as the fraction of stars that host such planets \citep{Jackson2012}.
Some stars also show evidence that could point to recent giant impacts
in the more distant reaches of their planetary systems \citep{Telesco2005,Dent2014,Stark2014}, but
conclusions remain tentative for now. 

The aim of this chapter is to review the topic of giant impacts, focussing on the potential of
observations of giant impact debris to inform on planet formation processes.
It starts with evidence for giant impacts in the Solar System (\S \ref{ss:giss}) before
discussing the various possible observable consequences of such impacts in extrasolar
systems (\S \ref{ss:observables}).
Generic aspects of the evolution of impact-generated debris are then discussed in
\S \ref{ss:gideb}.
The observability of debris from the Moon-forming giant impact is considered
in \S \ref{ss:moon} and then compared with observations of extrasolar debris disks
(\S \ref{ss:comparison}).
Giant impacts at larger distances from the star are discussed in \S \ref{ss:larger},
followed by a discussion of the implications from debris observations for the
formation of the ubiquitous super-Earth class of planetary system (\S \ref{ss:superearth}).
Conclusions are given in \S \ref{s:giconc}.

%% file: issigi_giss.tex
\section{Giant impacts in the Solar System}
\label{ss:giss}

In discussing the role of giant impacts in the formation of planets it is beneficial to turn first to 
our own Solar System, as here we have detailed information about the final end products -- the planets 
and satellites themselves.  Aside from Jupiter and possibly Saturn, true twins for the Solar System 
planets (in both size and orbital location) remain tantalisingly outside detection limits of searches 
for extrasolar planets at present.  While future missions will certainly extend these detection limits 
and allow us to begin detecting true Solar System twins, we will presumably always have more 
information about our own backyard than other star systems.  As is common in discussions of planet 
formation we divide our overview of the evidence for giant impacts in the Solar System into the inner 
Solar System (interior to Jupiter) and the outer Solar System (exterior to and including Jupiter).

\subsection{Inner Solar System}
\label{sss:innerss}

The inner Solar System is replete with evidence for giant impacts.  Each of the terrestrial planets 
has features for which a giant impact explanation has been suggested, with the exception of Venus, and 
it is probably no coincidence that Venus is also the terrestrial planet whose surface and geological 
history are least well understood.

\paragraph{Mercury}
\label{ssss:giss:mercury}

In comparison with the other terrestrial planets Mercury is unusually rich in iron, with a very large 
core. This extreme iron enrichment can be reproduced by a collision between a roughly 2.25~$M_{\rm 
Mercury}$ chondritic composition proto-Mercury and a roughly 0.4~$M_{\rm Mercury}$ impactor at speeds 
of around 30~km~s$^{-1}$ \citep[e.g.][]{Benz1988, Cameron1988, Anic2006, Benz2007}.  This impact speed 
is around 6 times the escape velocity of proto-Mercury and the resulting impact is extremely violent, 
almost destroying the planet.  While commonly thought of as a `mantle-stripping' event, in physical 
terms it is gravitational re-accumulation that is responsible for bringing the core of the former 
planet back together, rather than the outer layers being peeled away to reveal the exposed core. It is 
also notable that proto-Mercury in this scenario is very similar in mass to Mars, being around 16 per 
cent more massive.

Debris is a key component of the formation of Mercury via a giant impact, accounting for a larger 
fraction of the initial mass than the final planet.  As noted by \citet{Gladman2009} the fate of this 
debris, and in particular the fraction that is re-accreted, is essential to the final success or 
failure of a giant impact model to reproduce Mercury.

\paragraph{Earth-Moon system}
\label{ssss:giss:earthmoon}

Perhaps the most striking thing about Earth (from a dynamical perspective) is the large mass of the 
Moon.  At 1.2 per cent of Earth's mass the Moon is, by a substantial margin, the largest satellite 
relative to the size of its parent planet in the Solar System.

To find a comparable system one must look to the dwarf planet binary Pluto-Charon with a mass ratio of 
0.12, and which is also believed to be the result of a giant impact (\S~\ref{ssss:giss:plutocharon}).  
The angular momentum of the Earth-Moon system is also unusually high -- equivalent to a 4 hour 
rotation period in an isolated Earth, much faster than the other terrestrial planets.  Meanwhile 
tracing back the tidal evolution of the Moon suggests that it would have been much closer to Earth in 
the early Solar System, and compositionally the Moon is substantially depleted in both iron and 
volatile elements relative to Earth.

All of these factors led to the suggestion that the Moon might be the product of a giant impact 
between proto-Earth and another body \citep[e.g.][]{Cameron1976, Canup2004b}.  Continued work on the 
giant impact hypothesis eventually resulted in what has become known as the Canonical model 
\citep[e.g.][]{Canup2001, Canup2004a, Canup2004b}, in which a roughly Mars mass body (known as Theia) 
collides with a nearly fully formed proto-Earth at an oblique angle of $\sim45^{\circ}$ and a speed 
near the mutual escape velocity.  Such impacts would have been quite common in the late stages of 
terrestrial planet formation \citep[e.g.][]{Agnor1999}, and the Canonical model is very successful at 
reproducing many aspects of the Earth-Moon system.

In recent years the Canonical model has faced something of a crisis in the near indistinguishability 
of Earth and Moon in isotope ratios, which has spawned a re-think of the giant impact scenario 
\citep{Asphaug2014}. Some have invoked the evection resonance to permit the Moon to be created with 
collision parameters that are quite different to the Canonical model, satisfying both the dynamical 
and isotopic constraints \citep[e.g.][]{Canup2012, Cuk2012, Reufer2012}, while others have argued that 
the Canonical model parameters still satisfy the isotopic constraints, e.g., if some equilibriation of 
the accreted material occurs in the protolunar disk \citep{Pahlevan2007, Salmon2012} or if the 
impactor formed sufficiently close to the Earth \citep{MastrobuonoBattisti2015}, but the basic idea of 
the origin of the Moon in a giant impact does not seem to be in doubt.

\paragraph{Martian hemispheric dichotomy}
\label{ssss:giss:mars}

The largest topographic feature on Mars is the difference in elevation by $\sim$5~km between the lower 
northern and higher southern hemispheres \citep[e.g.][]{Smith1999} commonly referred to as the 
hemispheric dichotomy.  The possibility of the northern lowlands representing a massive impact basin 
was first suggested by \citet{wilhelms1984}. With more recent work \citep[e.g.][]{AndrewsHanna2008, 
Nimmo2008, Marinova2008, Marinova2011} this suggestion has become more concrete, and a giant impact 
appears to be the best explanation for the hemispheric dichotomy.

A giant impact is also suggested as a possible origin for the Martian moons Phobos and Deimos 
\citep[e.g.][]{Rosenblatt2011, Rosenblatt2012, Citron2015}.  The Borealis basin impact might thus be 
responsible for both the hemispheric dichotomy and the formation of the Martian moons.

\paragraph{Asteroid families and (4) Vesta}
\label{ssss:giss:vesta}

The Hirayama asteroid families are the result of large collisions within the asteroid belt 
\citep[e.g.][]{Durda2007}.  Though material strength begins to play an important role in asteroid 
sized bodies large impacts between asteroids share many of the features of giant impacts between 
planetary bodies \citep[e.g.][]{Jutzi2015}.  In the case of (4) Vesta and the associated Vesta family 
asteroids there is particularly strong link.  The second largest asteroid (mean diameter 525~km), 
Vesta is believed to be differentiated and possess a largely intact crust, and has been suggested as 
an intact proto-planet, or even `the smallest terrestrial planet' \citep[e.g.][]{Keil2002, 
Russell2012}.  In addition the numerous Vesta family asteroids can be traced to the formation of the 
massive Rheasilvia basin, which together with the underlying Veneneia basin dominate the southern 
hemisphere of Vesta.  Models of the formation of Rheasilvia indicate a $\sim$66~km diameter impactor 
\citep{Jutzi2013}.  The formation of the Rheasilvia basin can thus be thought of in some senses as a 
recent ($\lesssim$1~Gyr ago, \citealt[e.g.][]{Binzel1997}) `mini' giant impact.

\subsection{Outer Solar System}
\label{sss:outerss}

\paragraph{Pluto-Charon}
\label{ssss:giss:plutocharon}

As mentioned above the Pluto-Charon system is the closest analogue to the Earth-Moon system, despite 
lying on average over 40 times further from the Sun and having a substantially different composition, 
being rich in icy material.  Pluto-Charon is well described as a binary, having a mass ratio of around 
8.5:1, both bodies being tidally locked to one another, and with the system barycentre lying outside 
Pluto.  The angular momentum in the Pluto-Charon system is larger than could be contained in a single 
body, suggesting a collisional origin \citep[e.g.][]{McKinnon1989} and \citet{Canup2005} showed that a 
giant impact is indeed a good explanation for the Pluto-Charon binary.  The smaller satellites in the 
Pluto-Charon system may also have formed in the giant impact that produced Charon 
\citep[e.g.][]{Canup2011}, but explaining the complex resonant structure of the small satellites 
presents some difficulties \citep[e.g.][]{Showalter2015}.

\paragraph{Haumea collisional family}
\label{ssss:giss:haumea}

Analogous to the Hirayama families in the asteroid belt there is presently one collisional family 
known in the Kuiper belt, that associated with the dwarf planet Haumea \citep{Brown2007}.  This 
collisional family has a significantly lower velocity dispersion than the escape velocity of Haumea, 
and it has thus been suggested that the collisional family may have been created in an impact with a 
precursor of the current two satellites of Haumea \citep{Schlichting2009, Cuk2013} rather than Haumea 
itself.

\paragraph{Middle-sized moons of Saturn}
\label{ssss:giss:saturn}

It has been suggested that the formation of satellites around the gas giant planets is in many ways 
like a miniature version of the formation of the Solar System \citep{Canup2002, Canup2006}.  This 
would include giant impacts between proto-satellites, and such impacts may explain the origin of the 
otherwise mysterious middle-sized moons (those $\sim$500-1500~km in diameter) of Saturn 
\citep{Sekine2012, Asphaug2013}.  In this scenario Saturn would have started out with a system of 
larger satellites like the Galilean moons of Jupiter, but whereas in the Jovian system these moons 
remained locked in a stable Laplace resonance, in the Saturnian system they became unstable and 
collided, with the present middle-sized moons representing the remnants of this former satellite 
population.

\paragraph{Obliquity of Uranus}
\label{ssss:giss:uranus}

One of the important features of both the Earth-Moon and Pluto-Charon systems that is well explained 
by a giant impact scenario is their large angular momentum.  Given the large amount of angular 
momentum that can be imparted by a giant impact, if the impact orientation is substantially different 
to the spin axis of the planet it is possible for the impact to lead to a large change in the planet's 
obliquity.  It has long been recognised that an impact, involving an impactor around the size of 
Earth, represents a possible explanation for the large obliquity of Uranus 
\citep[e.g.][]{Safronov1972, Parisi1997}.
An impact is not the only means by which Uranus can be tilted;
tidal evolution of a massive retrograde satellite has also been proposed
\citep{Greenberg1974, KuboOka1995}, though
the moon required would have had to be greater than 1.5 times the mass of Mars and
thus within an order of magnitude of the mass of proposed impactors.

%% file: issigi_observables.tex
\section{Giant impact extrasolar observables}
\label{ss:observables}

One conclusion from \S \ref{ss:giss} is that there is abundant evidence for giant impacts
occurring in the Solar System.
Indeed, the majority of the planets show some evidence for giant impacts.
It is also the case that evidence for the impacts is manifested in many different ways,
in terms of the physical or dynamical properties of the planets and their environments.
The various consequences of giant impacts are summarised in Table~\ref{tab:observables},
along with the evidence in the Solar System for these outcomes and the potential for
observing them in extrasolar systems which is discussed further below.

\begin{table}[!htb]
\caption{Summary of the consequences of a planetary system being subjected to giant impacts:
the evidence for this from the Solar System and the possible manifestation in the observable
properties of extrasolar systems.}
\label{tab:observables}
\begin{tabular}{p{0.3\linewidth} p{0.3\linewidth} p{0.3\linewidth}}
\hline\noalign{\smallskip}
What do giant impacts do? & Evidence in Solar System & Potential extrasolar observable \\
\noalign{\smallskip}\hline\noalign{\smallskip}
Formation of moons & Earth-Moon, Pluto-Charon, Saturn? & Transit timing variations from exomoons \\
Modify internal composition & Mercury's Fe-rich composition & Exoplanet densities inferred from radial velocities and transits \\
Modify planet surface and atmosphere & Mars' hemispheric dichotomy, past magma ocean & Surface effects subtle, but hot planet spectrum \\
Modify planet spin & Uranus tilt & Exoplanet spins are measurable \\
Create debris & Hirayama asteroid families; dust bands; craters on small bodies & Infrared signature \\
\noalign{\smallskip}\hline
\end{tabular}
\end{table}

The formation of moons is one of the most obvious lines of evidence for giant impacts,
since this has emerged as the favoured explanation for our own Moon, with other Solar
System moons also explained in this way.
Given the small mass of any extrasolar moon, it seems like an impossible task to identify them. 
However, planets as small as $0.066M_\oplus$ have been detected in orbit around nearby stars
\citep{JontofHutter2015}.
The mechanism to find such low mass planets involves looking for small perturbations to the timing
of the transit of another large planet in front of the star due to the dynamical
perturbation of the smaller object (so-called transit timing variations, or {\it TTV}s).
This may be the most promising way of detecting extrasolar moons, although none are
known yet \citep{Kipping2014}.
The formation of a moon through giant impact requires an intermediate stage of a
circumplanetary disk.
There may be evidence for the passage of a circumplanetary ring system in front of its
host star 1SWASP J140747.93-394542.6 \citep{Mamajek2012, Kenworthy2015}.
While the nature of that object remains unsure, this may provide evidence for
satellite formation processes similar to those discussed in \S \ref{sss:outerss} for Saturn's moons.
The thermal emission from circumplanetary disks may also enhance the brightness of
an extrasolar planet, as has been suggested as the explanation for the spectrum of the
Fomalhaut-b planet \citep{Kalas2013}, and it is suggested that rings may also be
detectable in transit observations \citep{Zuluaga2015}.

A moon is not always an outcome of a giant impact, since in some cases very little material is
placed into orbit around the planet, one example being the loss of Mercury's mantle
in a violent collision to leave a high density planet (\S \ref{sss:innerss}).
The number of extrasolar planets for which densities have been measured has increased
dramatically recently due to the combination of high sensitivity transit observations
with Kepler (that measure the size of the planet) and radial velocity observations
(that determine the mass of the planet). 
In some cases the transit measurements also determine the mass of the planet, providing
there are nearby planets from which to measure the TTVs as mentioned above.
Such densities are used to infer the composition of the planets, with densities possibly
as high as 14\,g/cm$^3$ being recorded \citep{Marcy2014}.
This is a promising avenue, but one which has yet to be used to infer a collisional
history for the planets, rather this information has been used to assess the presence or absence
of an atmosphere.

The effect on the planetary surface is more subtle.
For example, the Martian hemispheric dichotomy is only apparent to us because
its surface height has been measured at high precision;
the level of difference is 0.07\% the diameter of Mars.
Variations in surface height may be measured from the shape of the transit of
an exoplanet, but the precision required is beyond current instrumentation.
However, if the collision is energetic enough, the surface would be melted in the
collision resulting in a magma ocean at 1000-3000\,K for rocky planets.
This molten surface would only be temporary since the heat would be radiated
away \citep[e.g., Earth's magma ocean may have survived 2\,Myr;][]{Zahnle2007},
but could have a dramatic effect on the emission spectrum of the
planet.
Whether the hot surface itself is observable depends on the planet's
atmospheric properties, which would in turn be modified by the impact;
e.g., the early nebula atmosphere could be stripped by impacts
\citep{Schlichting2015,Inamdar2015},
and the atmosphere could also be replenished by magma ocean outgassing. 
However, it is clear that the excess energy must be radiated somehow, and this
can increase the detectability of the planets following impacts
\citep{MillerRicci2009,Lupu2014}.
Such impact afterglows would be easier to detect at larger distance from
the star due to favourable contrast with the stellar emission.
This lead to the suggestion that the companion found at 55\,au from 2M1207 was in
fact the result of an impact between a $7M_\oplus$ and 74$M_\oplus$ planets
\citep{Mamajek2007}.
While more detailed analysis shows that the properties of this system are better 
explained by the atmospheric properties of a low mass companion without any
collision required \citep{Barman2011}, this at least demonstrates that impact
afterglows are in the realm of current observational capabilities.

Conservation of angular momentum requires a planet's spin to be modified as
a result of a giant impact.
In some cases this can be significant, perhaps explaining the tilt of Uranus'
spin relative to its orbital plane.
Exoplanet spins are already measurable in observations of the velocity profile from
lines in their atmospheres \citep{Snellen2014}.
For now such observations have been applied to one of the most massive known
extrasolar planets ($\sim 8M_{\rm jup}$ $\beta$ Pic-b), showing it to spin much
faster than its Solar System counterparts, likely due to its higher mass
as opposed to its collisional history.
There are also prospects to measure the orientation of an exoplanet's spin axis
with respect to its orbital plane using observations of its radial
velocity during secondary eclipse in a manner analogous to the Rossiter-McLaughlin
effect \citep{Nikolov2015}.

It is a general outcome of a collision that some fraction of the mass of the two
bodies does not remain bound to either one at the end of this process.
Indeed, the loss of Mercury's mantle in a giant collision not only modified
the composition of that planet, but also released large quantities of debris.
That debris, made up of solid material ranging in size from $\mu$m-sized 
dust grains to km-sized planetesimals, as well as vaporised rock if the collision was energetic 
enough, ended up orbiting the Sun.
There is abundant evidence for the creation of debris in collisions in the asteroid belt,
ranging from witnessing dust released in an impact which occurred 
in the last few years \citep{Jewitt2011}, to clustering in asteroid orbital elements showing these 
to have been created in an event a few Myr in the past \citep{Nesvorny2002}, evidence for which is 
also present in the dust bands in the inner Solar System \citep{Grogan2001}, to the 
Hirayama asteroid families which may have been much more violent events Gyr-ago 
\citep{Hirayama1918}.
Evidence for debris released in the giant impacts proposed to have shaped 
the Solar System's planets is harder to find since these collisions likely occurred long ago, 
shortly after the Solar System formed.
Nevertheless such evidence may exist in the cratering record and chemical composition
of asteroids and meteorites \citep{Bottke2015}.

From an extrasolar perspective, giant impact debris is expected to be readily detectable,
because even relatively small quantities of dust (e.g., that corresponding to an asteroid
of a few 10s of km in diameter ground into $\mu$m-sized dust) can be detected around nearby stars due
to its thermal emission.
The dust is heated by the star and then re-radiates that emission in the infrared at levels
that can exceed the star's own photospheric flux \citep{Wyatt2008}.
Some 20\% of nearby stars show evidence for circumstellar dust originating in the break-up
of planetesimals.
While a large fraction of this dust is likely produced in the steady
state grinding down of extrasolar analogues to the Kuiper belt, since such
an interpretation explains the statistics of disk detection as a function of age
\citep{Wyatt2007b}, some fraction of these may instead be the signature of a recent giant
impact, since giant impact debris is expected to emit at potentially detectable
levels \citep[e.g.,][]{Jackson2012}.

To summarise, there are many ways in which evidence for giant impacts may be
found in extrasolar systems.
While there has been much progress in identifying these signatures, for now 
the most promising evidence for giant impacts in extrasolar systems lies 
in the creation of debris, and it is to this aspect that the remainder of this chapter
will be devoted.

%% file: issigi_gideb.tex
\section{Evolution of debris from giant impacts}
\label{ss:gideb}

As noted in \S \ref{ss:observables}, one of the key observable features produced by a giant impact around 
another star is debris.
Though the character of the observable features will vary somewhat depending on 
the nature of the impact, for example the mass of the colliding bodies, or the orbital distance at
which the impact occurs, the underlying processes which shape these features and how
they evolve possess many similarities.  
Here we briefly outline a general framework within which the evolution of giant impact debris
may be understood \citep[for a more detailed discussion see][]{Jackson2014}.

\subsection{Dynamics of giant impact debris}
\label{sss:gideb:dynamics}

\begin{figure}[!htb]
\centering
 \includegraphics[width=0.8\textwidth]{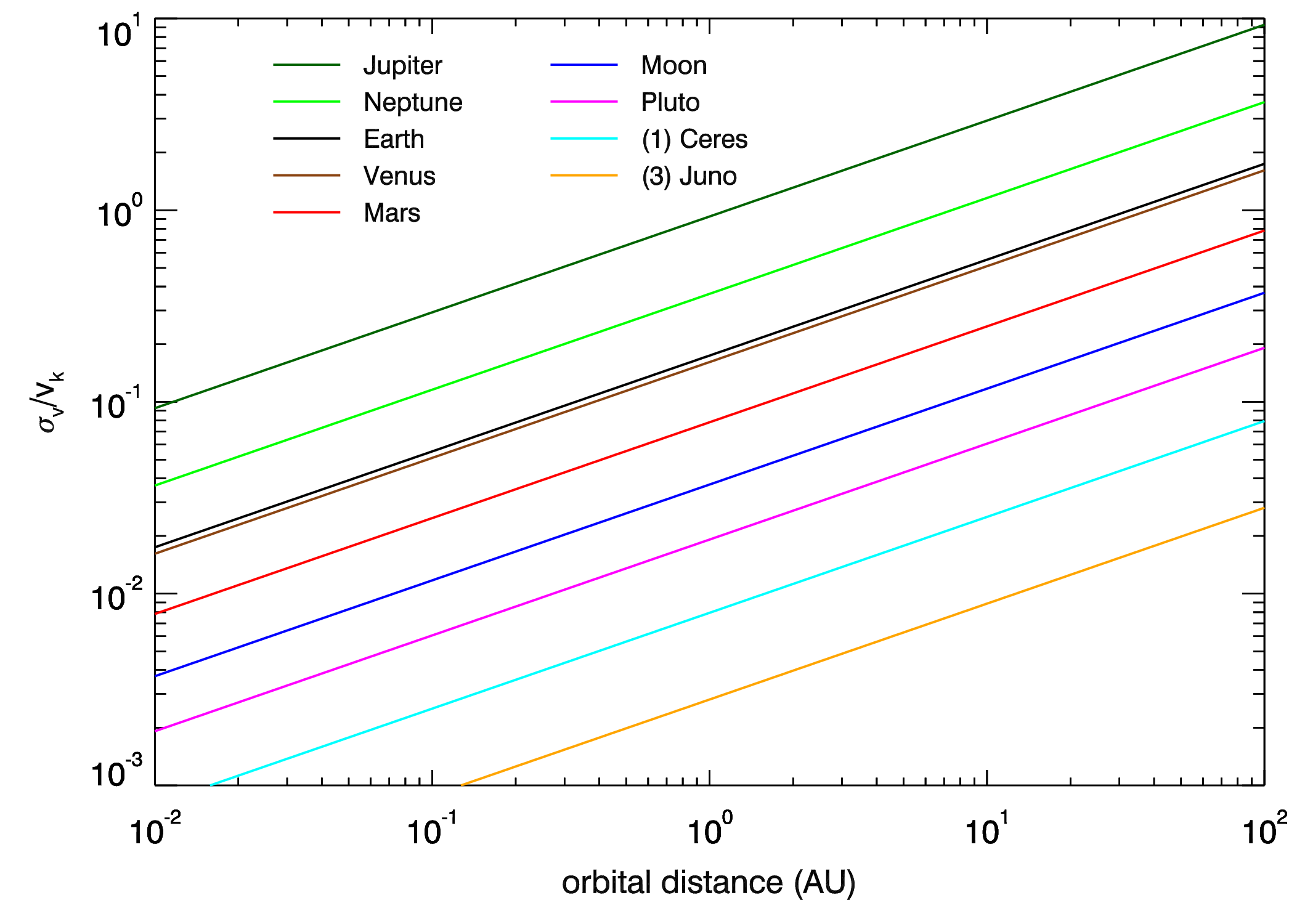}
 \caption{The scaled velocity dispersion imparted to giant impact debris, $\sigma_v/v_{\rm k}$,
  as a function of orbital distance of the progenitor for a selection of Solar System bodies
  ranging in mass from Neptune to the asteroid (3) Juno,
  under the assumption that $\sigma_v = 0.46 v_{\rm esc}$.
  This ratio is important for the morphology and subsequent evolution of the debris;
  e.g., much of the debris may be placed onto orbits that are unbound from the star
  if $\sigma_v/v_{\rm k} \gg 1$.}
 \label{fig:sigvvk-orbdist}
\end{figure}

The dynamical evolution of debris released in a giant impact can be considered as beginning from a single 
point with velocities relative to the star given by the progenitor's Keplerian orbital velocity $v_{\rm k}$ 
plus some velocity dispersion.
While the velocity dispersion is likely to be asymmetric, it is convenient to model this as isotropic 
given that the orientation of any asymmetry would be specific to a given impact, and is likely to be 
random.
The velocity dispersion of debris in giant impact simulations is approximately Gaussian with a width of 
$\sigma_v$ \citep[e.g.][]{Jackson2012}, where $\sigma_v$ is comparable to the
escape velocity of the progenitor $v_{\rm esc}$.
The subsequent dynamical evolution and morphology of the debris depends on the ratio $\sigma_v/v_{\rm k}$ which 
is shown in Fig.~\ref{fig:sigvvk-orbdist} for different progenitors as a function of orbital 
distance, using the assumption that $\sigma_v \approx 0.46v_{\rm esc}$ as found for
the Moon-forming impact \citep{Jackson2012}.
Note that the velocities of the debris being comparable to the escape velocity of the
progenitor (regardless of the progenitor mass) holds largely independently of the
velocity distribution of the debris particles.
This is because to go into heliocentric orbit the debris must have sufficient
energy to escape the progenitor body, and once the energy required to escape
is exceeded the resulting relative speed post-escape then rapidly becomes
comparable to the escape velocity.
For example a particle that exceeds the escape velocity by only 5\% at
launch will then have a relative speed post-escape of $0.32v_{\rm esc}$.

To understand the evolution and morphology of the debris it is important to remember that immediately 
after the impact the debris occupies what is essentially, on the scale of the wider planetary system, a 
single point at the location of the impact, which we will refer to as the {\it collision-point} 
\citep{Jackson2012,Jackson2014}.
This geometry means that the orbits of all debris fragments must pass through this point, which thus 
has a high density.
A small density enhancement is also found on the opposite side of the star from the collision-point, 
since all debris orbits must pass through the plane of the progenitors' orbit somewhere along
what we will refer to as the {\it anti-collision line} \citep{Jackson2012, Jackson2014}.
Since debris passes through the collision-point once per orbit, the resulting repeat encounters can
also have implications for the rate at which debris encounters the progenitor and indeed other debris. 

\begin{figure}[!tb]
 \includegraphics[width=0.49\textwidth]{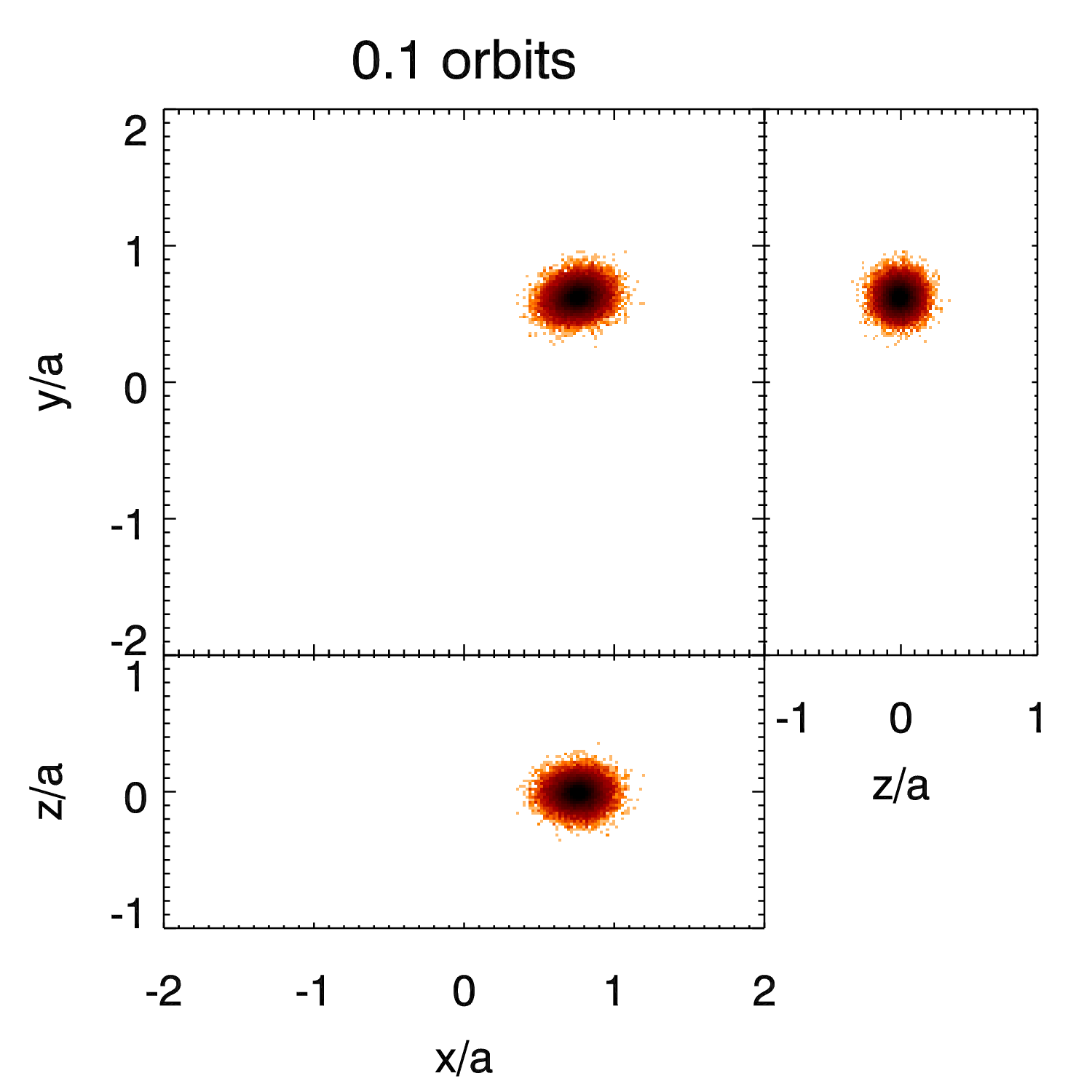}
 \includegraphics[width=0.49\textwidth]{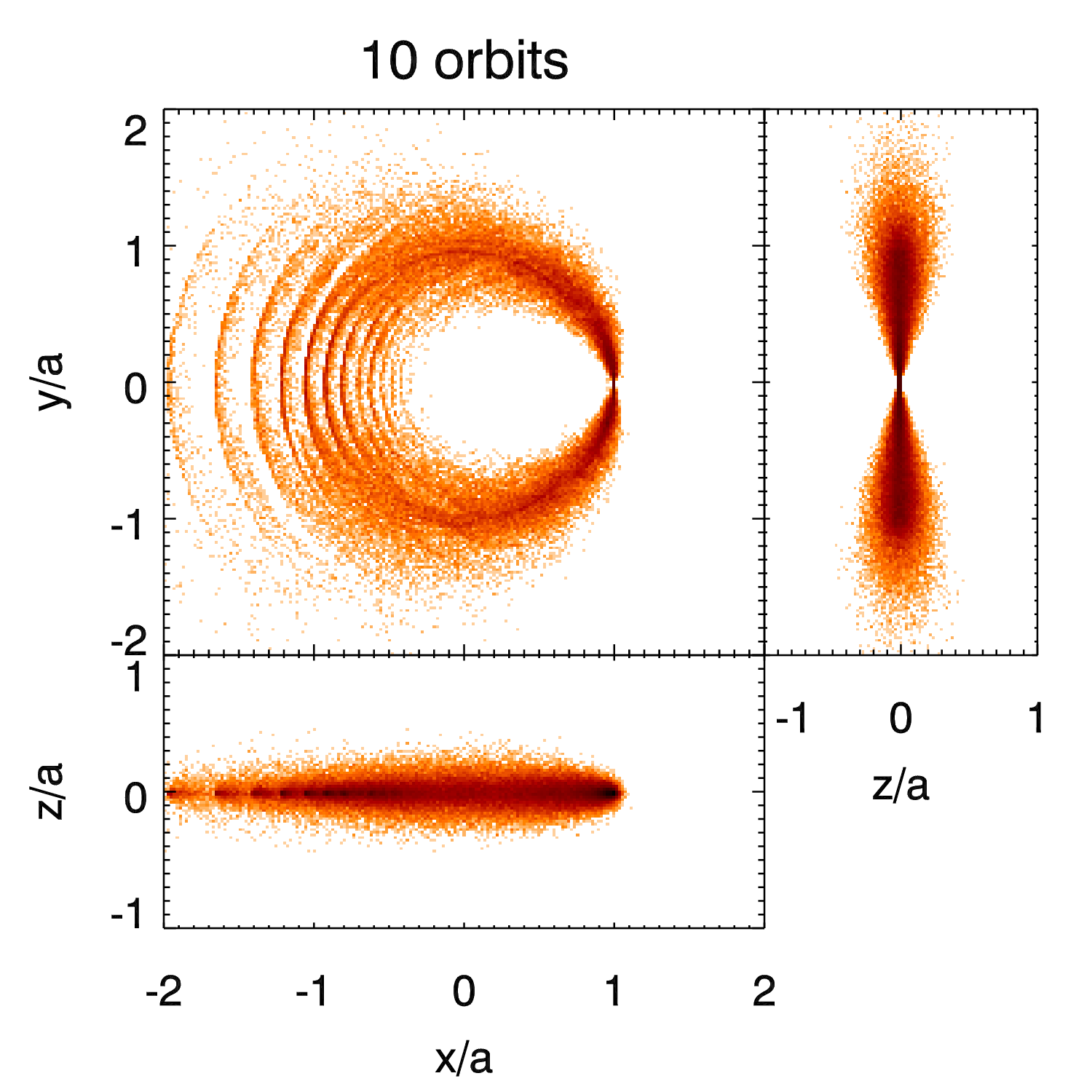}
 \includegraphics[width=0.49\textwidth]{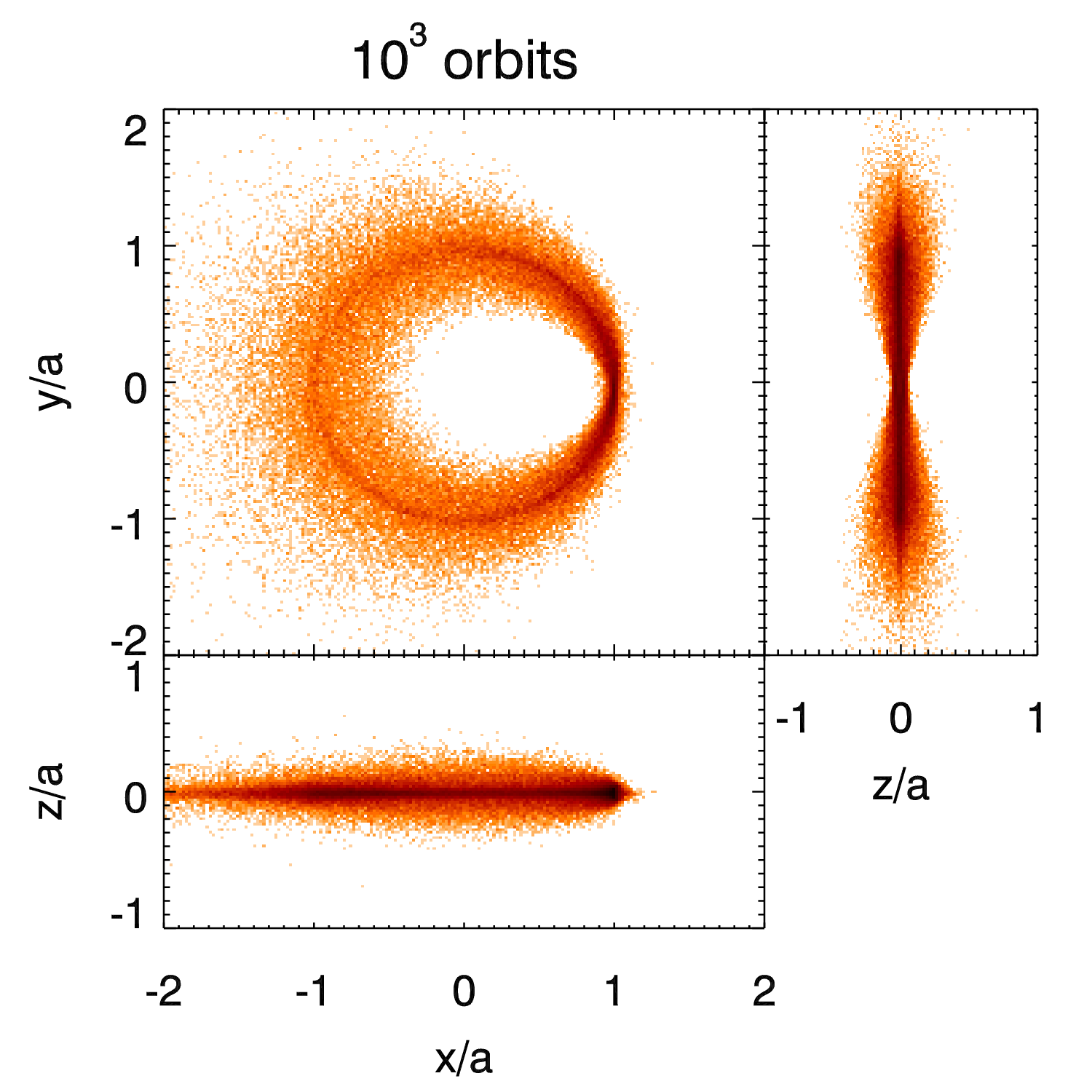}
 \includegraphics[width=0.49\textwidth]{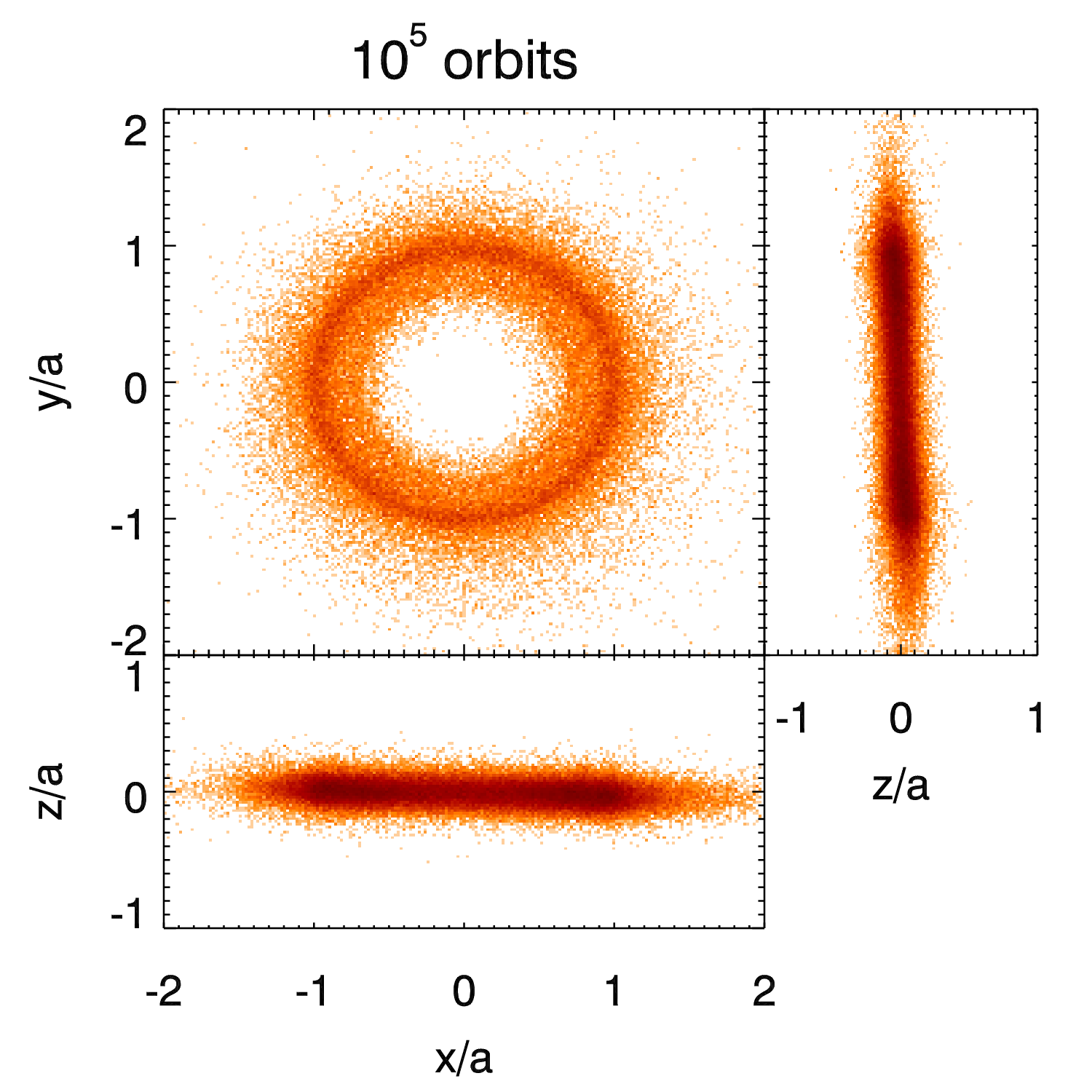}
 \includegraphics[width=\textwidth]{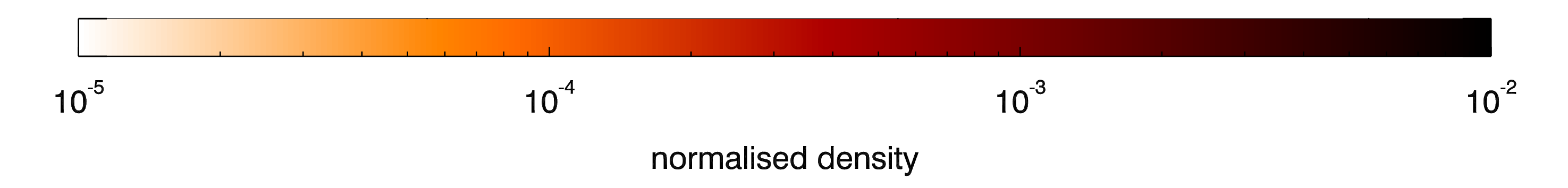}
 \caption{Snapshots of the distribution of debris ejected by a giant impact with $\sigma_v/v_{\rm k}=0.18$,
  showing the four key stages of its dynamical evolution: (top left) clump,
  (top right) spiral, (bottom left) asymmetric disk and
  (bottom right) axisymmetric disk.
  For each stage the top left panel shows a face-on view of the disk with the star
  at the origin and the collision-point at (1,0), while the panels to the right and below
  this show edge-on views.
  The axis scale is in units of the progenitor semi-major axis, $a$, while the snapshot times are in units of the
  progenitor's orbital period.
  The colour corresponds to the disk density normalised to an integrated value of 1 at time zero.}
 \label{fig:moondynsnaps}
\end{figure}

Common to the dynamical evolution of all (bound) giant impact debris are four key 
morphological phases:
{\bf (i)} After the impact, both the debris and the progenitor move away from the collision-point.
The debris initially forms an expanding clump that moves with the progenitor body around its orbit
(see upper left panel of Fig.~\ref{fig:moondynsnaps}).
{\bf (ii)} Since fragments that have been placed onto orbits with smaller semi-major axes than the
progenitor will have shorter orbital periods than those with larger semi-major axes, the clump will
be sheared as it continues around the orbit.
Once the shearing has progressed enough that fragments on smaller orbits have reached the
collision-point before those on larger orbits have left it, the clump will become a spiral
structure, which then continues to coil as Keplerian shearing continues (see upper right
panel of Fig.~\ref{fig:moondynsnaps}).
{\bf (iii)} Eventually the spiral structure of the debris is no longer visible and the
debris forms a smooth but highly asymmetric disk, with a strong pinch at the collision-point
(see bottom left panel of Fig.~\ref{fig:moondynsnaps}).
The density enhancement at the collision-point, as well as how spread out the disk
is away from the collision-point, both increase with velocity dispersion
\citep[i.e., with $\sigma_v/v_{\rm k}$, for more detail see][]{Jackson2014}.
{\bf (iv)} The orbits of debris fragments precess due to interactions with massive bodies,
including the progenitor, at a rate that depends on the semi-major axis of the fragments.
The resulting differential precession eventually smears out the asymmetry producing an axisymmetric disk
(see bottom right panel of Fig.~\ref{fig:moondynsnaps}).

The timescale for the shearing process, and so for the evolution from clump to asymmetric disk,
depends on $\sigma_v/v_{\rm k}$.
The shearing is quite rapid in Fig.~\ref{fig:moondynsnaps} since $\sigma_v/v_{\rm k}=0.18$;
the early spiral stage is visible after 1 progenitor orbit, and after 100 orbits the spiral 
is no longer distinguishable.
In the case of an impact with a smaller $\sigma_v/v_{\rm k}$ the rate of Keplerian shearing is slower,
since the velocity difference across the debris distribution is smaller.
While this would result in slightly longer clump and spiral phases, in general the clump phase is
expected to end within ten orbits, and the spiral phase within a few hundred orbits after the impact.
Since the morphology of the disk, and the
rate at which the disk progresses through the early phases of the evolution, depend on $\sigma_v/v_{\rm k}$,
this means that impacts involving progenitors of the same mass can have substantially different
morphologies and different early stage evolution if they occur at different orbital distances (see
Fig.~\ref{fig:sigvvk-orbdist}).

The smooth asymmetric phase lasts much longer, since it is determined by the precession timescale, rather
than by the dynamical timescales that set the clump and spiral phases.
Precession rates depend on the system architecture, i.e., the masses and orbits of
other planets in the system, although the progenitor itself may also be massive
enough to set the precession period.
In general the collision-point may be expected to have been smeared out after a few thousand
orbits and the disk to have achieved total axisymmetry after a few tens of thousands of orbits,
timescales that apply to precession due to a Jupiter mass planet 5 times more distant than the
impact \citep[see eq. 23 of][]{Jackson2014}, or due to an Earth-mass progenitor \citep{Jackson2012}.

The morphological evolution shown in Fig.~\ref{fig:moondynsnaps} will occur whatever the wider
architecture of the planetary system in which the progenitor and debris reside.
However, the progenitor and other planets play a potentially much more important role beyond
setting the precession timescale of the debris, since close debris-planet approaches can lead to the
debris being accreted onto the planet or ejected from the system.
Which of these outcomes is more likely depends on the ratio of the escape velocity of the planet
to its Keplerian velocity, $v_{\rm esc}/v_{\rm k}$:
ejection is more likely for $v_{\rm esc}/v_{\rm k} \gg 1$ and accretion more likely for $v_{\rm esc}/v_{\rm k} \ll 1$
(see Fig.~\ref{fig:sigvvk-orbdist}).
The rate at which debris is accreted onto a planet of radius $R_{\rm p}$
can be estimated as
\begin{equation}
  R_{\rm col} = n \sigma_{\rm col} v_{\rm rel},
\label{eq:racc}
\end{equation}
where $n$ is the density of debris in the vicinity of the planet, 
$v_{\rm rel}$ is the relative velocity of encounters between the planet and the debris,
and $\sigma_{\rm col} = \pi R_{\rm p}^2 [1 + (v_{\rm esc}/v_{\rm rel})^2]$ is the collision
cross-section.\footnote{Note that the gravitational focussing factor for re-accretion onto the progenitor is
independent of the progenitor's properties, since $v_{\rm rel}$ scales with $\sigma_v$ and so $v_{\rm esc}$.
This factor is found to be close to 10 for the Moon-forming giant impact debris.}
However, note that this rate may be significantly underestimated at early times for re-accretion onto
the progenitor if the geometrical effect of the collision-point is not taken into account;
during the early phases of dynamical evolution, most re-accretion onto the progenitor will occur when
it passes through the collision-point, and the accretion rate will be substantially raised during 
this early period \citep[][see also discussion in \S \ref{sss:gideb:collisions}]{Jackson2012}.

\subsection{Collisional evolution}
\label{sss:gideb:collisions}

In addition to evolving through dynamical interactions with massive bodies in the system, the debris will 
also evolve as a result of mutual collisions within the debris population.
Such collisions lead to the break-up of larger fragments into smaller ones,
redistributing mass down the debris size distribution.
Eventually the debris is ground down into dust grains that are small enough to be
removed by radiation pressure on orbital timescales.
The size at which this occurs is called the blow-out size and for dust of density $\rho$
is given approximately by \citep[e.g.][]{Wyatt2008}
\begin{equation}
D_{\rm bl} = 0.8 \left(\frac{L_*}{L_{\odot}}\right) \left(\frac{M_{\odot}}{M_*}\right)
             \left(\frac{\rm 2700\hspace{1ex} kg\hspace{1ex} m^{-3}}{\rho}\right)
             \hspace{1ex}\mu{\rm m}.
\label{eq:dbl}
\end{equation}

The rate at which debris fragments collide with each other is
determined by an equation that is analogous to eq.~\ref{eq:racc}.
This calculation is complicated by the fact that any given debris particle will
interact with particles with a range of sizes, with the outcomes of such collisions ranging
from cratering to complete pulverisation.
For this reason the collisional evolution of debris populations is often studied numerically
\citep{Thebault2003,Krivov2005,Bottke2005}.
However, this is not necessary to understand what happens, because for most common
assumptions about collisional outcomes, once the size distribution has reached steady state it
tends to a shape that can be readily calculated \citep{OBrien2003,Wyatt2011}.
If it can be assumed that the strength of a debris fragment is independent of its
size, the size distribution tends to a power law
\begin{equation}
  n(D) \propto D^{-\alpha},
  \label{eq:nd}
\end{equation}
where $\alpha=7/2$.
Such a size distribution has most of the mass in the largest objects, while most of its
cross-sectional area is in the smallest objects (i.e., those close to the blow-out limit
in eq.~\ref{eq:dbl}).

It is important to understand how both mass and cross-sectional area evolve within the debris
distribution.
It is the cross-sectional area to which the observations discussed in \S \ref{ss:observables}
are sensitive \citep{Wyatt2008}.
However that cross-section resides in small grains that are relatively short-lived, and so must be
replenished from the larger debris fragments which make up the majority of the mass and
have longer collisional lifetimes.
The discussion is simplified if the size distribution has a fixed shape, because
this implies that cross-sectional area scales with mass by a constant ratio, though
this is only true at all times if the debris size distribution starts with the size
distribution it will tend to in steady state.
Typically the total mass of debris created in a giant impact is well known.
For example, for large bodies (those with $\sigma_v/v_{\rm k}>0.1$), impacts
generally fall in the hit-and-run or partial accretion/erosion regimes with debris
releases of around $3-5$\% of the colliding mass \citep{Leinhardt2012, Stewart2012}.
This is because for large $\sigma_v/v_{\rm k}$, the impact speed must be a
larger fraction of the orbital speed to push the impact into a violent outcome
state.
Smaller bodies (those with $\sigma_v/v_{\rm k} \ll 0.1$) are more likely
to undergo violent impacts where debris masses can be a much higher fraction of the
colliding mass, but if the impact parameters are fairly well constrained the
debris mass can also be determined with reasonable confidence.
On the other hand the size distribution is not well constrained, as discussed further in
\S \ref{sss:brightness}.
This is unfortunate, as for a fixed mass of giant impact debris, both its cross-sectional
area and (as we will see below) the timescale of its evolution, are strongly dependent on the
size of object where most of the mass is created.

Nevertheless, the above discussion motivates a simple prescription with which to predict the
evolution of cross-sectional area of the giant impact debris.
It is only necessary to consider the evolution of the mass, which will deplete due to
the dynamical processes discussed in \S \ref{sss:gideb:dynamics},
and due to the collisional erosion of the largest fragments.
A quick estimate for the rate at which the largest fragments deplete by collisions
can be made using eq.~\ref{eq:racc} and assuming that all the debris
is in fragments of the same size.
While more accurate calculations are readily available
\citep[see][]{Wyatt2007a}, this illustrates how the collisional
depletion rate is proportional to the mass of largest fragments.
This leads to the mass remaining constant until the largest objects have come
to collisional equilibrium, following which the mass decays inversely with time, with
dynamical processes further accelerating the depletion process.

\begin{figure}[!tb]
\centering
  \includegraphics[width=0.8\textwidth]{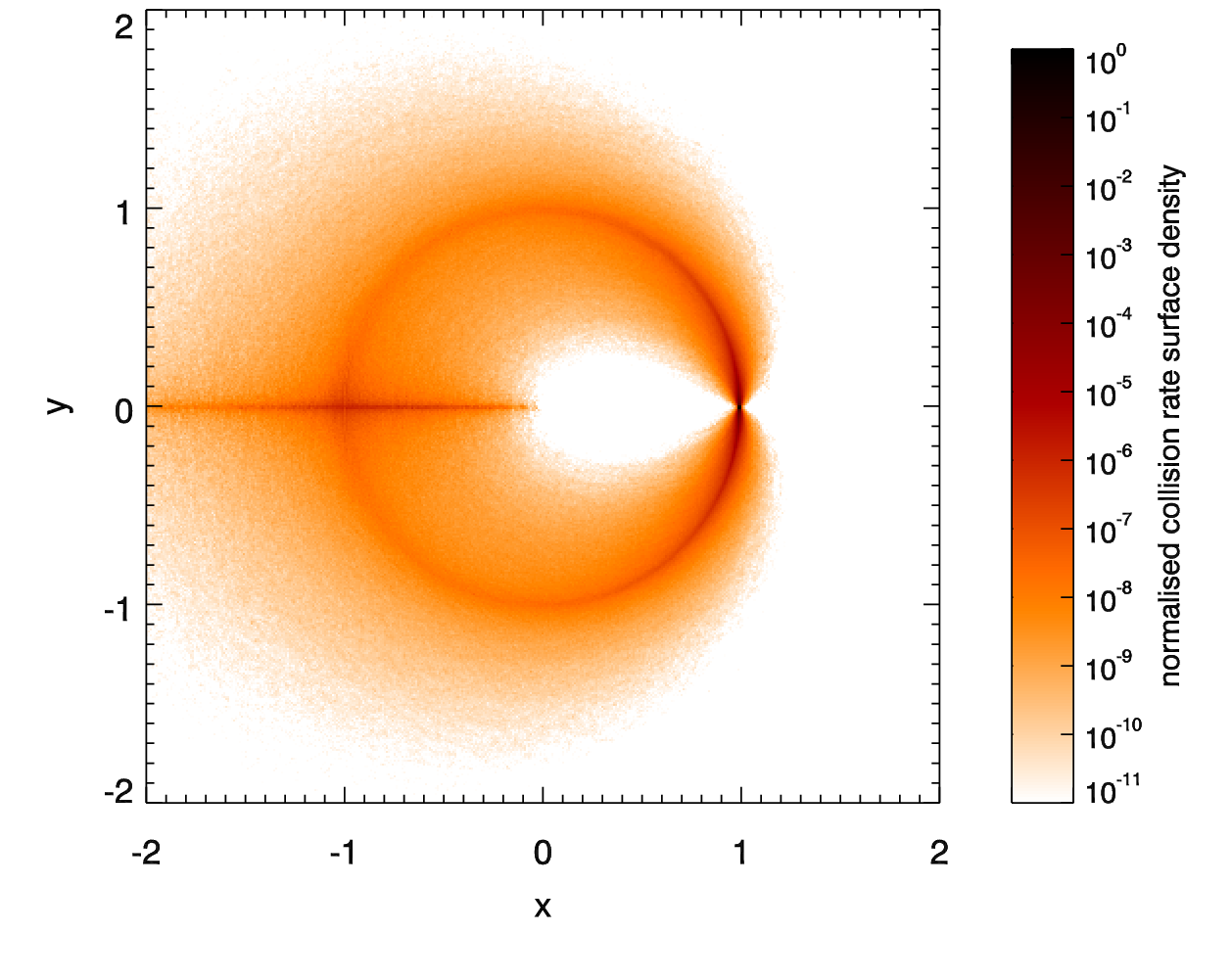}
  \caption{Image showing how the rate of mutual debris collisions varies with position during
    the asymmetric phase of evolution of the debris produced by a giant impact with
    $\sigma_v/v_{\rm k}=0.3$.
    The rates are normalised such that the maximum collision rate is 1.
    Orbital motion is in the anti-clockwise sense.
    The geometry of the debris orbits results in an increased density along
    $y=0$ (see Fig.~\ref{fig:moondynsnaps}), and so a higher collision rate at
    the collision point and along the anti-collision line. }
\label{fig:collmap}
\end{figure}

There are several caveats to the above picture.
For example, it is known that the strength of a debris fragment is dependent on its size,
and the debris size distribution may not be expected to start with the shape it tends
to in steady state.
Both are valid criticisms, but their consequences can be readily accounted for
\citep{Lohne2008,Wyatt2011}.
Furthermore, it is found that the simple prescription given above is usually accurate
to within a factor of a few and so is acceptable for many applications.

A more complicated caveat to address is the geometrical effect of the collision-point.
In the early phases of the dynamical evolution of the disk
(prior to axisymmetrisation) the number density varies substantially around the orbits of
the debris fragments;
in particular the collision-point is a region of exceptionally high density.
Fig.~\ref{fig:collmap} shows how this results in the collision-point having a collision rate
that is many orders of magnitude higher than elsewhere.
This effect is such that, to a first approximation, all collisions that a debris particle experiences
occur close to the collision-point.
It is also such that the collision rate averaged over the particle's orbit is around two orders of 
magnitude higher than would be the case for an axisymmetric disk;
i.e., the rate of collisional evolution will be raised during the early phases of
dynamical evolution.
The density enhancement along the anti-collision line also leads to a slightly increased
collision rate there, though this is much less dramatic.
These geometrical effects can only be accounted for using numerical techniques \citep[e.g.,][]{Kral2015},
and have further implications which will be discussed in \S \ref{ss:larger}.

%% file: issigi_moon.tex
\section{Observability of debris from the Moon-forming giant impact}
\label{ss:moon}

Though recent work has put the Canonical model in some doubt (see \S \ref{sss:innerss}),
the Moon-forming giant impact remains the most well studied example of a giant impact,
and thus provides a good starting point for discussions of the behaviour and observability
of debris released by giant impacts \citep[for more detail see][]{Jackson2012}.

\subsection{Dynamical evolution}
\label{sss:dynamical}

From smoothed-particle hydrodynamics simulations conducted by \citet{Marcus2009} the velocity 
dispersion of the debris released by the Moon-forming giant impact is well fit by a Gaussian with a 
width of $\sigma_v=5.2$\,km~s$^{-1}$.
At an orbital distance of 1~au this corresponds to $\sigma_v/v_{\rm k}=0.18$ 
(the value used in Fig.~\ref{fig:moondynsnaps}).
Once it has been launched from Earth the debris goes into heliocentric orbit and the debris
distribution then evolves dynamically as described in \S~\ref{sss:gideb:dynamics}, with the
proviso that the general discussion therein only gave brief consideration to planets in
the system other than the progenitor, which were not included in Fig.~\ref{fig:moondynsnaps}.

\begin{figure}[!tb]
  \centering
  \includegraphics[width=0.8\textwidth]{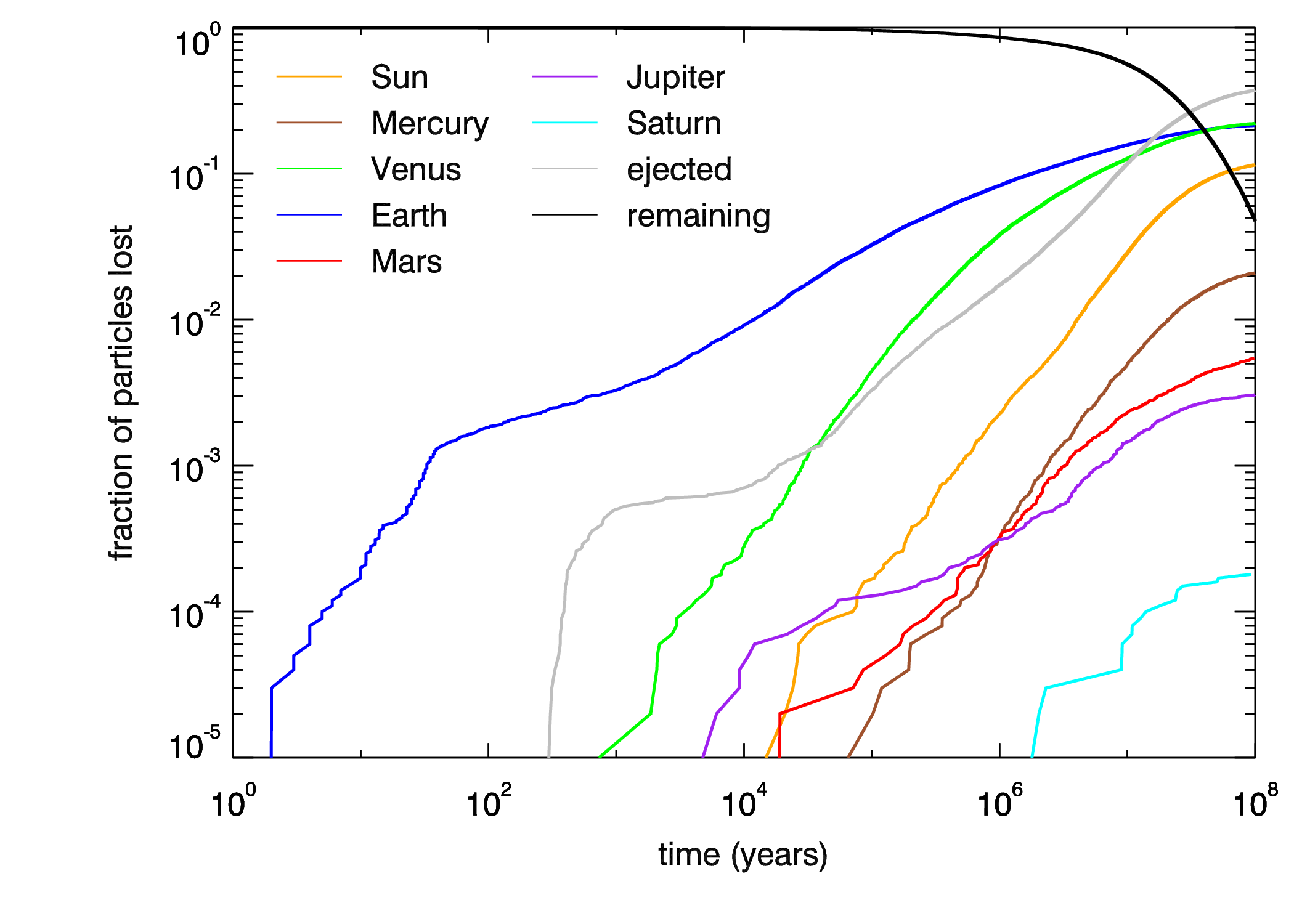}
  \caption{The fates of particles in a 100\,Myr 10$^5$ particle $N$-body simulation of
  debris from the Moon-forming giant impact.
  Debris particles are massless test particles, and no account is made for collisional erosion.
  The present day Solar System architecture is used.
  Lines for different planets indicate the fraction of particles accreted onto that planet.}
  \label{fig:moondebacc}
\end{figure}

The wider architecture of the planetary system becomes important as the giant impact debris
leaves the asymmetric stage.
The two most important planets aside from Earth in determining the dynamical evolution of the debris 
are Venus and Jupiter.
Venus draws the debris inwards such that, rather than being narrowly peaked around the orbit
of Earth, the debris occupies a broad band encompassing the orbits of both Earth and Venus.
Jupiter meanwhile truncates the outer edge of the disk by ejecting fragments that encounter 
it from the Solar System.
The differing effects of Venus and Jupiter are in line with expectations from
Fig.~\ref{fig:sigvvk-orbdist}, since Venus has $v_{\rm esc}/v_{\rm k} \ll 1$ while
Jupiter has $v_{\rm esc}/v_{\rm k} \gg 1$.
Fig.~\ref{fig:moondebacc} shows the ultimate fate of debris created in the Moon-forming
impact, which shows that the three fates of accretion by Earth, accretion by
Venus, and ejection (mostly by Jupiter) account for the overwhelming majority of the
dynamical losses of the debris.
The next most likely fate, collision with the Sun, is in fact also a secondary effect of Jupiter;
in systems lacking Jupiter the quantity of debris which impacts the Sun is much lower.
In total half of the dynamical particles have been lost after 13\,Myr, while just 5\%
of the dynamical particles remain after 100\,Myr.

While of course other planetary systems may have very different orbital architectures to the Solar 
System, we can expect that some general features of the evolution of the debris from Moon-formation 
will apply to all systems.
In particular we can expect that the most important other bodies in the system will be nearby
terrestrial planets of comparable (or larger) mass than the progenitor, and the nearest giant planet.
More distant giant planets have limited influence, since the nearest one will almost always
eject particles before they have a chance to interact with more distant giant planets
(i.e., as long as $v_{\rm esc}/v_{\rm k} \gg 1$).

\subsection{Predicted brightness evolution}
\label{sss:brightness}

\begin{figure}[!tb]
  \includegraphics[width=\textwidth]{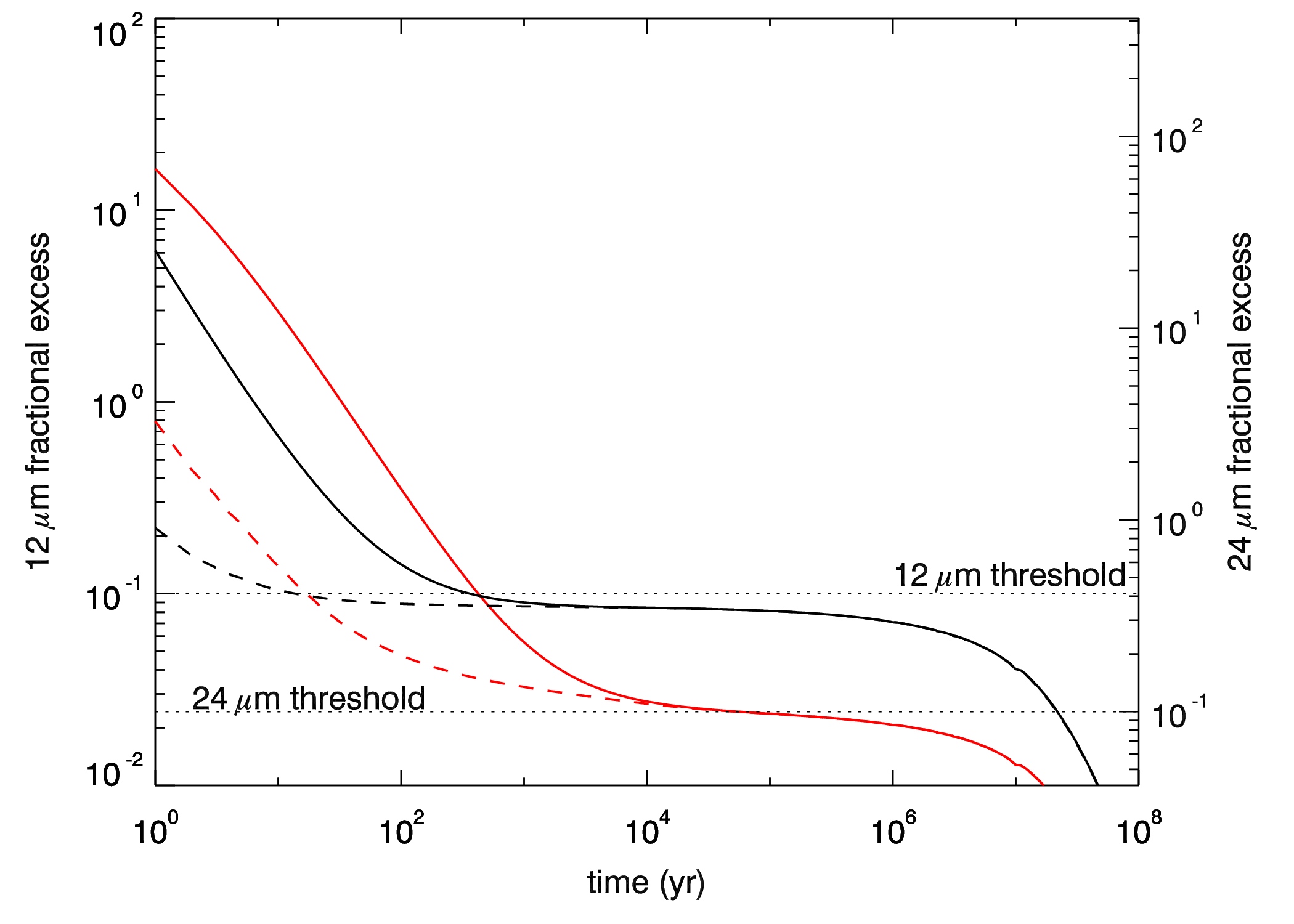}
  \caption{Evolution of the fractional excess of debris released by a Moon-forming impact
  around a Sun-like star for different assumptions about its initial size distribution and
  about its collisional evolution.
  The fractional excess is defined as the flux from the debris relative to the flux
  from the stellar photosphere, shown at 12\,$\mu$m on the left axis and 24\,$\mu$m
  on the right axis;
  the dotted lines show typical detection limits at fractional excesses of 10\%
  at both wavelengths.
  The black lines show the evolution of a two component size distribution with
  30\% vapour condensates up to 1~cm in size and 70\% boulders up to 500~km in size.
  The red lines show the evolution of a three component size distribution with 30\% vapour
  condensates up to 1~cm in size, 50\% intermediate material up to 7.5~km in size
  and 20\% large boulders up to 500~km in size.
  Solid lines calculate collision rates assuming the debris to be axisymmetric at all times,
  while dashed lines show the results incorporating some aspects of the collision-point
  geometry in this calculation (see text for details).}
  \label{fig:EMbrightevol}
\end{figure}

As discussed in \S~\ref{sss:gideb:collisions}, the brightness of giant impact debris depends
on the quantity of small dust present, which in turn depends on how the debris size
distribution evolves due to collisions.
Several issues with calculating the collisional evolution were discussed in
\S~\ref{sss:gideb:collisions}, where it was noted that there may be inherent
uncertainties arising from how well the initial size distribution is known,
and most collision rate calculations also suffer from an inaccurate treatment
of the geometry of the collision-point.
For this reason Fig.~\ref{fig:EMbrightevol}, which shows the predicted brightness
evolution of giant impact debris at two wavelengths 12 and 24~$\mu$m at which stars
are commonly observed, shows that evolution for two different initial size
distributions and two different treatments of the collisional evolution.

There are additional uncertainties involved in translating dust
cross-sectional area into the emergent emission spectrum.
In all cases on Fig.~\ref{fig:EMbrightevol} the emission from the dust is calculated
assuming the grains to act like simple blackbodies at the orbital distance of the progenitor.
This has the benefit of resulting in a constant ratio between the 12 and 24~$\mu$m
emission so that both can be shown on the same plot.
However, the small grains that dominate the emission are expected to be hotter than black body,
while at the same time emitting inefficiently, and potentially exhibiting spectral features
due to their composition (e.g., \S \ref{sss:composition}).
These effects can be properly accounted for, but require additional assumptions (e.g., about
the grain composition).
For simplicity these are not included in Fig.~\ref{fig:EMbrightevol}, which
should thus be considered to have additional factors of a few or more uncertainty.

As mentioned in \S~\ref{sss:gideb:collisions} the size distribution of debris
released by a giant impact is poorly constrained.
SPH simulations of giant impacts are unable to resolve anything but the 
largest debris fragments \citep[$\gg$100~km for Earth-mass impactors;][]{Genda2015}.
The main constraint from such simulations is the total mass of the debris, which is around
0.016~$M_{\oplus}$ ($\sim 10^{23}$~kg) in the Canonical model.
However, SPH simulations of giant impacts do indicate that, due to the enormous energy
involved in the collision of two planetary-sized bodies, a significant fraction of the
debris material is released in the form of vapour.
As the cloud of vapour expands it cools and condenses, forming small droplets with typical
sizes in the mm-cm range \citep[e.g.,][]{Melosh1991, Johnson2012, Johnson2014}.
For impacts involving Earth-sized bodies, typically a few tens of percent of the debris 
is released in the form of vapour;
\citet{Canup2008} estimated $\sim$10-30\% for the Canonical 
Moon-forming impact, though \citet{Kraus2012} suggest that the equations of
state in use in current hydrodynamical codes may underestimate vapour production.
The production of vapour alongside material that is not vaporised suggests that the size
distribution should consist of (at least) two components, the vapour condensates and
unvaporised {\it boulder} material that extends to larger sizes.

The simplest assumption for the initial boulder size distribution is that it starts
with the shape it would tend to in steady state (e.g., eq.~\ref{eq:nd}) and
extends up to some maximum size $D_{\rm{max}}$.
This was the assumption made in \citet{Jackson2012}, which is reproduced in
Fig.~\ref{fig:EMbrightevol} for $D_{\rm{max}}=500$\,km with the solid black line.
This illustrates the properties common to all models for the brightness evolution.
In the early phases the vapour condensates shine brightly at readily detectable levels.
However, these are rapidly depleted by collisional evolution to non-detectable levels within
a few hundred years.
Dust arising from the break-up of the boulders starts at a lower level, but is maintained
for much longer, eventually starting to deplete after a few Myr due to a combination of
collisional erosion and dynamical erosion (see Fig.~\ref{fig:moondebacc}).
Taken at face value, the boulder distribution remains undetectable throughout
the evolution, and the total emission from both vapour condensates and boulders
drops below the typical 12\,$\mu$m detection threshold after a few hundred years.
However, the predicted excess level is so close to the typical detection threshold
that, given the various factor of a few uncertainties mentioned elsewhere, 
the giant impact debris could be detectable at 12\,$\mu$m up to 10\,Myr,
which is its duration of detectability at 24\,$\mu$m.

An alternative choice for the initial size distribution of the boulder material is that of 
\citet{Bottke2015}, basing it instead on the size distribution of the Vesta family asteroids, with 
additional constraints from Lunar cratering.
The size distribution of the Vesta family is considerably steeper than a self-similar cascade, 
with more mass in smaller bodies.
\footnote{It is unclear how appropriate it is to extrapolate the Vesta family size
distribution to planetary scale giant impacts.
For example, the Rheasilvia impact speed of $\sim 15 v_{\rm esc}$ is much faster than typical
giant impact speeds, while at $\sim$0.1\% of the mass of Vesta the 66~km impactor is
unusually small \citep{Jutzi2013}.
The largest of the Vesta family asteroids (aside from (4) Vesta itself) is only 
$\sim$8~km in size, and gravitational re-accumulation may become more efficient at larger sizes.  
Additionally, the size distribution of Vesta family asteroids may have undergone
collisional evolution in the $\sim$1~Gyr since the Rheasilvia impact.
Nevertheless, it may be the only observable example of a giant impact-like debris distribution.}
\citet{Bottke2015} suggest that the boulder material may itself have two components,
one a steep power law ($\alpha \sim 5$ in eq.~\ref{eq:nd}) extending to a largest size
of $\sim$80-100~km and an additional less massive population of larger bodies.
The red lines in Fig.~\ref{fig:EMbrightevol} show the predicted brightness evolution for a
three-component size distribution intended to mimic that of \citet{Bottke2015}.  
The resulting evolution is very similar to that of the two-component distribution
considered in \citet{Jackson2012}.
In particular, it is noticeable that the brightness evolution exhibits just two rather
than three distinct phases (i.e., a drop over $\sim 100$ years followed by another
after $\sim 10$\,Myr), which is because the emission from the grinding of
intermediate-sized ($\sim 10$~km) material overlaps with the peak from the vapour condensates
(since this population is both bright and has a very short collisional lifetime).
The relatively low mass in the largest boulders with this assumption
makes it less likely that the debris remains detectable up to 10\,Myr after the
impact event, but any such conclusion would bear the caveat of the factors of few
uncertainty in brightness.

The solid lines on Fig.~\ref{fig:EMbrightevol} made the same assumption as \citet{Jackson2012}
in calculating the collision rate in assuming the debris to be distributed in an axisymmetric
torus.
The assumption of axisymmetry is common in calculations of collision rates
\citep{Wetherill1967,Wyatt2010}, since precession causes orbits to precess on long timescales.
However, in the early phases of evolution the orbital planes and pericentres have yet
to be randomised due to precession, and moreover in the clump phase it is incorrect
to assume that particles are evenly distributed along their orbits.
While such effects are automatically accounted for in the N-body simulations for accretion
of debris onto planets, these simulations do not follow debris-debris collisions due to
the large number of particles that would be required.
It is possible to make a more accurate calculation of the collision rates using the
output of the N-body simulations.
The dashed lines on Fig.~\ref{fig:EMbrightevol} show the evolution using collision rates 
calculated from the output of the N-body simulations in which the orbital planes and
pericentre orientations are no longer assumed to be random
\citep[see Fig.~\ref{fig:collmap} and][]{Jackson2014}.
As expected from the discussion in \S \ref{sss:gideb:collisions}, the decay in
brightness occurs much faster due to the geometry of the collision-point.
However, there is a caveat in that this calculation still assumed that particles are evenly
spread around their orbits.
This means that the collision rate is overestimated during the clump phase, and the true
evolution is likely to lie somewhere between the solid and dashed lines on
Fig.~\ref{fig:EMbrightevol}.

A further caveat that applies during the earliest clump and spiral phases is the prospect
for the debris to have high optical depth. 
All of the brightness calculations discussed so far assumed the debris to be optically thin,
whereas in fact the dust arising from the break-up of vapour condensates is predicted to be
so abundant that its distribution may be optically thick.
This could have several consequences, by changing the detectable cross-sectional area,
the dust temperature, and whether it is removed from the system by radiation pressure. 
These effects would also vary on orbital timescales.
Clearly further work is needed to improve our understanding of the evolution of giant
impact debris during the first tens of orbits.
Progress has been made in incorporating collisional evolution into $N$-body simulations
of the evolution of giant impact debris \citep{Kral2015}, however the small volume of
the collision-point means that it remains a challenge to accurately model its effect
due to the spatial resolution required to do so.

%% file: issigi_comparison.tex
\section{Comparison with observed systems}
\label{ss:comparison}

\subsection{Photometric fluxes}
\label{sss:photometric}

From \S \ref{sss:brightness} it is clear that if we were observing another star
shortly after its growing terrestrial planets had undergone an impact
similar to that which created the Earth-Moon system, then this event
would have had a significant effect on the level of infrared emission observed. 
Moreover, with multiple such events occurring, it seems that the Solar System
would have shone brightly in the mid-IR throughout its infancy.
Given the other evidence for collisions occurring throughout the history of
the Solar System (see \S \ref{ss:giss}), \citet{Rieke2005} suggested that the scatter
in levels of 24\,$\mu$m emission seen in their Spitzer survey of nearby A stars
(i.e., stars slightly more massive and luminous than the Sun), and their higher levels
in the first 150\,Myr of the stars' lives, could be a result of
recent collisions.
In that interpretation, high levels of mid-IR emission are attained
stochastically following a giant impact, with levels fading back to quiescent
levels thereafter.
A similar interpretation of 24\,$\mu$m emission levels was applied to a Spitzer
survey of Sun-like stars \citep{Meyer2008}, wherein it was assumed that this
emission is indicative of ongoing terrestrial planet formation processes.

However, the situation is complicated, because 24\,$\mu$m emission can
arise for a number of reasons unconnected with collisions or planet formation.
It was later shown that both the scatter seen in the \citet{Rieke2005} observations
and the decay in emission levels could be explained by the steady state erosion
of extrasolar Kuiper belts \citep{Wyatt2007b};
the level observed simply reflects the initial mass in the belt and its distance
from the star, which was far enough that the emission was generally cold 
and so also explained the observations at 70\,$\mu$m toward the same stars
\citep{Su2006}.
The same issue affected the interpretation of Sun-like stars, since it was
found that the emission spectrum for the majority of these stars was rising
toward longer wavelengths implying that the 24\,$\mu$m emission arises in a belt far enough from
the star ($>10$\,au) to be explained by steady state processes \citep{Carpenter2009}.
That is, while recent giant impacts are not ruled out as the origin of the
24\,$\mu$m emission, a steady state interpretation is more likely for most
systems, and regardless the relatively cool temperature of the emission implies that this
arises from a location outside that typically associated with the formation of terrestrial
planets.

The problem is simply that while 24\,$\mu$m emission should accompany a giant
impact (see Fig.~\ref{fig:EMbrightevol}), it is a flawed proxy for such an impact.
Nevertheless, the large samples of young stars observed by Spitzer mean that
the evolution of a star's (or more accurately a population of stars') 24\,$\mu$m
emission is well characterised \citep[e.g.,][]{Siegler2007}, and as such has often been
used to assess the frequency of giant impacts \citep{Jackson2012}.
More recently it has become clear that a better proxy for such collisions is
a star's excess 12\,$\mu$m emission.
Such hot emission is seen less frequently and cannot arise from a more distant
Kuiper belt, rather the requisite temperature to emit at such short wavelengths
places the dust at a temperature compatible with the location of the terrestrial
planets in the Solar System.
This does not preclude the presence of a more distant cold dust belt in the system,
and indeed there are many examples of debris disks with both cold outer belts
and hot inner dust \citep{Su2013,Kennedy2014};
the hot dust in some of the younger such systems has been attributed to ongoing
planet formation processes \citep{Smith2009}.

Early searches for 12\,$\mu$m emission from dust around nearby stars with IRAS were
not particularly successful \citep{Aumann1991}, though several good candidates
were rediscovered in the last decade and followed up with higher resolution instruments
to confirm the emission was associated with the star and to characterise its spectrum
\citep[e.g.,][]{Rhee2008,Melis2010}.
These hot excesses were picked out on a case-by-case basis, and interpreted
again within the context of terrestrial planet formation, based on the fact that
the stars with bright emission were predominantly found around $<100$\,Myr stars.

A significant advance in the statistical analysis of the frequency of stars
exhibiting excess 12\,$\mu$m emission came with WISE, which surveyed the whole
sky at that wavelength with a spatial resolution and sensitivity far superior to IRAS.
This enabled \citet{Kennedy2013} to quantify the rarity of bright excesses
by correlating Sun-like stars in the Hipparcos database with WISE.
They found that detectable excesses (those for which 12\,$\mu$m dust emission
is at levels above $\sim 10$\% that of the stellar photosphere) occur around
1:1000 stars.
While it is not feasible to derive the ages of all Hipparcos stars, an age
dependence for the excess emission was evident because the majority of the 22 stars
with excess were known to be members of young associations.
To make a rough estimate of the age dependence, \citet{Kennedy2013} made the
assumption that stars in the Hipparcos catalogue have ages that are randomly distributed
up to the main sequence age for their spectral type.
This led to the conclusion that for stars in the 10-120\,Myr age range, the frequency
of detectable emission is closer to 3\%;
i.e., the excess rate declines with age, just as it does at other wavelengths.

Eventually the WISE data should be mined to determine the excess distributions in
all young clusters, similar to the analysis carried out for Spitzer.
For now, this has been applied to a few regions such as 10-20\,Myr Upper Sco \citep{Luhman2012}
and 3-5\,Myr Taurus \citep{Esplin2014}.
The interpretation is complicated by the presence of long-lived protoplanetary disk systems
among the excess candidates.
However, based on this work we estimate the fraction of Sun-like stars
in Upper Sco with detectable 12\,$\mu$m excesses is $\sim 10$\%,
in broad agreement with the findings of \citet{Kennedy2013},
particularly if one considers that the fraction is likely to decrease over the 10-120\,Myr
age range (and noting that some of the Upper Sco excesses are from protoplanetary disks).

\begin{figure}
   \includegraphics[width=0.9\textwidth]{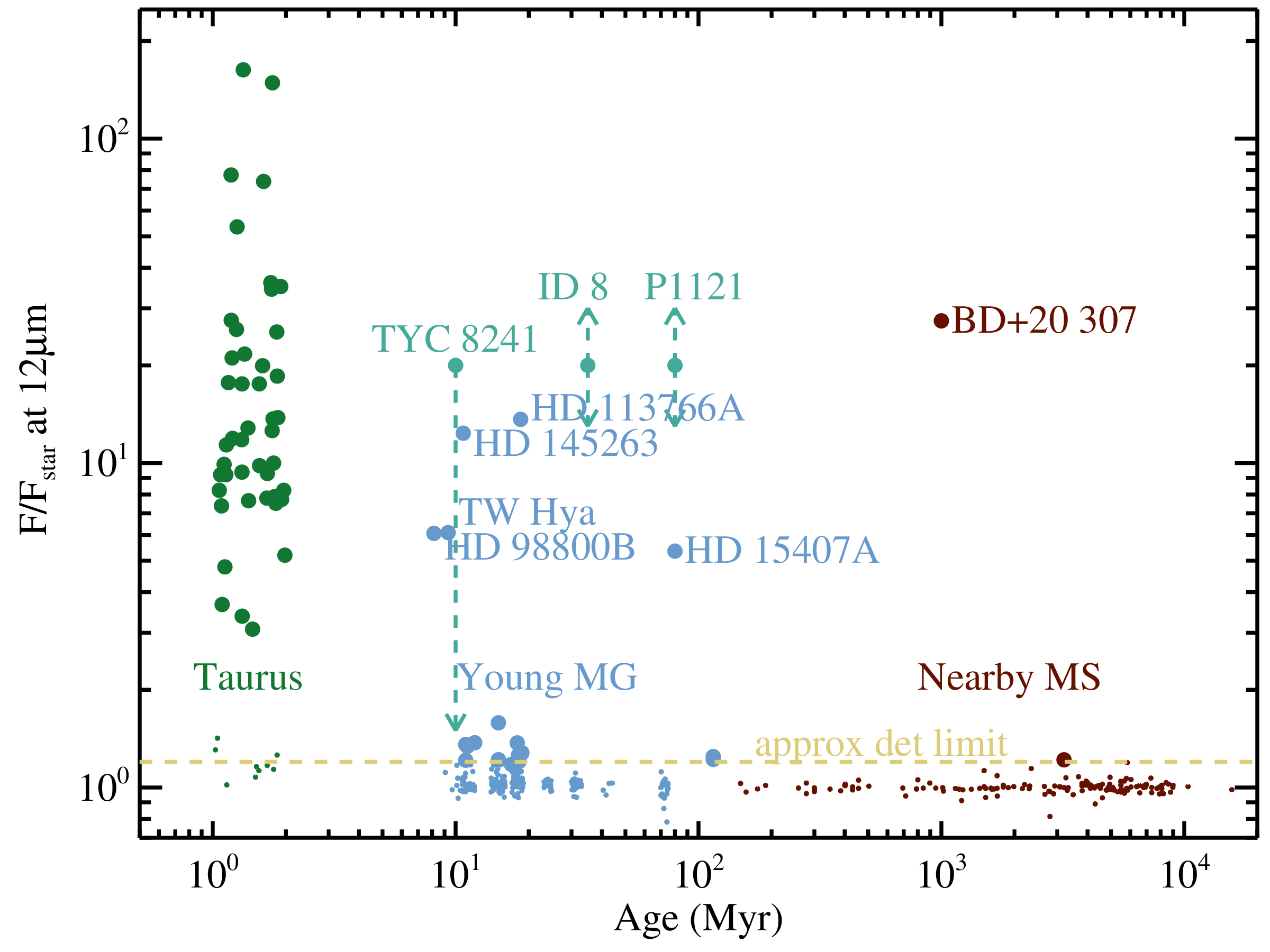}
  \caption{Evolution of 12\,$\mu$m emission around Sun-like stars.
  Large circles are those for which an excess above photospheric levels are detected,
  while small circles are those for which no excess is discernable;
  the approximate detection threshold is given by the horizontal yellow line, though this varies
  from star to star.
  The different colours indicate different surveys:
  green is Taurus \citep{Esplin2014},
  light blue are young moving groups such as Sco Cen \citep{Rizzuto2012}
  and our SED fitting of bona fide members from \citet{Malo2013},
  brown are our SED fits of nearby members from the DEBRIS sample \citep{Phillips2010}.
  Individually identified 12\,$\mu$m excess sources are also included from
  \citet{Kennedy2013}.
  The vertical dashed lines show the level of variability for sources discussed in
  \S \ref{sss:variability} \citep{Melis2012, Meng2014, Meng2015}.
}
\label{fig:r12}
\end{figure}

The 12\,$\mu$m excesses are plotted on Fig.~\ref{fig:r12} as a function of stellar age
for a number of Sun-like stars.
Some of these were identified in statistical samples from which the rate of excesses was
determined above, but others were simply excess detections reported in the literature such
that care should be taken with the interpretation of this plot. 
Nevertheless, this illustrates the level at which excess is detected, and how that level
is typically higher around younger stars.
In fact the level of emission around main sequence stars can reach the level seen around
T Tauri stars, meaning that the full emission spectrum needs to be considered when
interpreting a 12\,$\mu$m excess, as it is not always clear whether this should be
interpreted as evidence for a recent collision or for a protoplanetary disk
\citep{Schneider2013, Kennedyetal2014}.

\subsection{Compositional constraints}
\label{sss:composition}

The information provided by a detection of excess emission extends beyond the photometric
measurement of the level of excess.
For those systems with detections, the mid-IR spectrum provides information about the composition
of the dust \citep[e.g.,][]{Chen2006}.
A prime example of this is the 12\,Myr A5V star HD172555 at 29\,pc for which the mid-IR spectrum
is dominated by emission from amorphous silica dust \citep{Lisse2009}.
Dust of different compositions leaves signatures in the spectrum at specific wavelengths;
e.g., the silica feature at $\sim 9.3$\,$\mu$m allows its presence to be picked out by eye
for HD172555.
However, extracting more detailed information about the dust composition is not trivial,
since the spectral features are relatively broad so that those from different compositions
overlap, and the exact shape of the spectrum depends on the dust size
distribution and the temperatures attained by different sizes.

Debris disk modellers commonly employ Mie theory for interpretation of debris disk
observations \citep[e.g.,][]{Lebreton2013}, however the approximations in this approach
mean that such detailed spectra are often interpreted using optical properties obtained
from laboratory measurements of astrophysical materials \citep{Lisse2009}.
In the case of HD172555 it was the latter approach which was used, resulting in
the conclusion that around half of the dust is composed of silica, with a
total dust mass equivalent to a 150-200\,km asteroid.
Since silica is produced in high energy impacts, this points to an origin for
the dust in a recent collision at $>10$\,km/s between two massive protoplanets
each at least the mass of Mercury.
This interpretation was supported by a feature in the spectrum at 8\,$\mu$m.
While this was originally believed to indicate the presence of SiO gas created
in the collision \citep{Lisse2009}, the short lifetime of such gas due to
photodissociation suggested an alternative origin for the feature in solid SiO
created by condensing vaporised silicate \citep{Johnson2012}.
In summary, the composition is similar to that which would have been anticipated following
the Earth-Moon forming collision.
This analogy is reinforced by the dust temperature which places it at a distance of
$\sim 6$\,au from the star, in keeping with spatial constraints from imaging and
interferometry \citep{Smith2012}, a location that is comparable in terms of incident energy
to the  Solar System's terrestrial planets given the higher luminosity of this star.

One puzzling aspect of the dust orbiting HD172555 is its inferred size distribution, since
the prominence of the silica feature requires there to be abundant sub-$\mu$m dust,
whereas such dust is small enough that it would be expected to be blown away from the
star by radiation pressure on orbital timescales of a few years (see eq.~\ref{eq:dbl}).
This is in contrast to the persistence of the mid-IR emission at the same level to
within 4\% over 27 years (from IRAS to WISE), suggesting that the dust is created more
rapidly than it is removed, or that radiation pressure is somehow ineffective.
\citet{Johnson2012} preferred the latter explanation, and found that for certain
compositions it is possible for this star to retain a long-lived population of
$\ll 0.1$\,$\mu$m grains even if those $0.1-1$\,$\mu$m are significantly depleted;
this is because the effect of radiation pressure is strongest for $0.1-1$\,$\mu$m
grains, and for certain compositions such grains can have $\beta>1$ (and so must be
unbound) while smaller (and larger) grains have $\beta<0.5$ (and so can be bound).
However, the origin of the size distribution remains unclear, and may be related to the
issues discussed in \S \ref{sss:variability}, and may also affect the assumed detectability
of giant impact debris discussed in \S \ref{sss:brightness}.
Another relevant aspect of this system that we are just beginning to learn about
is its gas content;
for now it is known that OI is present \citep{RiviereMarichalar2012}, but its location
relative to the hot silica dust is not yet known, as are its origin and impact on the
hot dust.

While this section has focussed on just one system, there are many other systems
for which compositional constraints exist, not all of which exhibit the same
silica composition, even if they are interpreted within the framework
of giant collisions \citep{Lisse2008,Currie2011,Ballering2014}.

\subsection{Time variability}
\label{sss:variability}

Whereas the hot dust emission toward HD172555 was concluded to not be variable
(see \S \ref{sss:composition}), the same is not true for all stars with hot
dust mid-IR emission.
For example, that of BD+20307 brightened by 7\% over the course of the 6 months
between two WISE observations \citep{Kennedy2013}.
The case of TYC8241 is particularly interesting, because the emission persisted
at a high level of excess ($\sim 10$ times brighter than the star at 12\,$\mu$m)
for around 25 years before dropping by an order of magnitude to a negligible
level over the course of 2 years \citep{Melis2012}.
A good explanation for the disappearance of the hot dust is lacking.
One of the ideas proposed is that the dust was removed in a collisional
avalanche, a runaway process wherein the input of large quantities
of small grains that get blown out by radiation pressure
break-up larger debris on their way out \citep{Grigorieva2007}.
If so, the previously high levels of excess will return once the
population of small dust is replenished in collisions between the bigger debris
that was unaffected by the avalanche.

However, given the discussion in \S \ref{sss:composition}, perhaps it is not
the disappearance of the dust which is surprising, rather it is the earlier
persistence of the dust in the face of the star's radiation pressure which is more
surprising.
Indeed, a two year timescale for the emission to decline is compatible with the
orbital removal timescale expected for radiation pressure. 
The question then would be why radiation pressure was ineffective for so long, and
what then caused it to turn on so suddenly?

Another case that is particularly illuminating is that of 35\,Myr-old Sun-like (G6V)
star ID8 for which the 24\,$\mu$m emission was noted to be variable 
in \citet{Meng2012}. 
Later observations showed that the near-IR (3-5\,$\mu$m) emission to this star underwent
a substantial brightening followed by decay on a year timescale, with quasi-periodic
modulations (with period $\sim 30$\,days) in the disk flux \citep{Meng2014}.
This was interpreted as the result of a giant collision in which silicate vapour is
created out of which small particles condense and subsequently collisionally deplete.
The geometry of the collision-point during the early phases of the debris
evolution explains the periodic modulation (see \S \ref{sss:dynamical}),
implying that the collision was located at $\sim 0.33$\,au.
Subsequent monitoring of 5 other systems with large levels of mid-IR emission showed
that 5/6 of the full sample show significant variations on timescales less than a year
\citep{Meng2015}.

Regardless of the origin of this variability, there is an empirical implication
for how bright mid-IR emission is interpreted, since its level does not evolve in
the manner that the colder dust in debris disks is thought to evolve.
That is, the evolution is more complicated than that outlined in \S
\ref{sss:gideb:collisions} and \S \ref{sss:brightness}, in which the
dust emission was assumed to decay slowly as the largest objects are ground into
dust that is removed by radiation pressure (eq.~\ref{eq:dbl}).
Indeed, the shape of what \citet{Kennedy2013} called the exozodi luminosity
function, that is the fraction of stars with 12\,$\mu$m excess levels above a given
level, may already bear witness to that;
the luminosity function is relatively flat compared with the expectation that
the brightest known excesses will decay due to collisions such that their
brightness scales inversely with time.

\subsection{How common are late giant impacts?}
\label{sss:common}

In \citet{Jackson2012} the predicted evolution of 24\,$\mu$m emission from the debris
following the Moon-forming impact was compared with the statistics for nearby stars
to estimate the fraction of stars that form terrestrial planets.
This section carries out a similar calculation, but for the reasons described in \S \ref{sss:photometric}
is based on the evolution of the 12\,$\mu$m emission.
The question is also framed slightly differently, so that we seek to determine the
fraction of stars that undergo late giant impacts
\begin{equation}
  f_{\rm{gi}} = f_{12} * f_{\rm{12gi}} / f_{\rm{12gidet}}.
  \label{eq:fgi}
\end{equation}
That is simply the fraction of 10-120\,Myr stars with bright 12\,$\mu$m excess, $f_{12}$,
multiplied by the fraction of those bright excesses that originate in giant impacts rather
than bright asteroid belts, $f_{\rm{12gi}}$, divided by the fraction of the 10-120\,Myr timeframe
that it would be expected the giant impact debris would be at a detectable level, $f_{\rm{12gidet}}$.

One of these parameters is well constrained, since \S \ref{sss:photometric} showed that
$f_{12} \approx 3$\%, though this may be around a factor of 3 higher at the youngest end of
the age range.
While $f_{\rm{12gi}}$ is not well known, it can be said to be constrained to be less than
100\%.
One might be mistaken for thinking that this is close to 100\% from the 
literature, since mid-IR emission is usually interpreted as the result of giant impacts,
but there are not many cases where it is possible to rule out that the dust is simply the
bottom end of the collisional cascade of a massive asteroid belt, at least for systems
that are in this age range for which collisional processes have yet to have a significant
effect on depleting any remnant asteroid belt \citep{Wyatt2007a}.

The hardest parameter to estimate is $f_{\rm{12gidet}}$.
The most pessimistic assumption would be that there is just one giant impact following the
dispersal of the protoplanetary disk (i.e., the Moon-forming impact), and that the dust created
from the grinding of boulders is below the detection threshold. 
This would result in $f_{\rm{12gidet}} \approx 10^{-5}$ (i.e., the impact debris is detectable for
around 1000 years out of the $\sim 100$\,Myr timeframe).
In this case, the only way to explain the high detection rate of 12\,$\mu$m excesses would be for
the majority of these to originate in asteroid belts rather than giant impacts
(i.e., $f_{\rm{12gi}}<3\times10^{-4}$).
However, the conclusion is quite different if the dust from boulders starts out bright enough to be
detectable, since the debris from a single impact would remain detectable for $\sim 10$\,Myr,
and so $f_{\rm{12gidet}} \approx 0.1$.
This would mean that $f_{\rm{gi}}<0.3$.
This calculation starts to become more constraining if one considers that terrestrial planet formation
simulations predict that multiple giant impacts occur in the 10-100\,Myr timeframe.
Indeed, it is usually assumed that $f_{\rm{12gidet}} = 0.5-1$ \citep{Kenyon2004,Jackson2012,Genda2015}.
If this is the case, then the fraction of systems that undergo a late giant impact must be
less than 6\%.
This would provide a useful constraint for planet formation models, by indicating that either the
formation of Solar System-like terrestrial planets is relatively rare, or that planet growth is
usually complete by the time the protoplanetary disk disperses;
i.e., terrestrial planet formation is not normally accompanied by a late Moon-forming impact,
which means planet formation models need to be revised.
However, given the caveats on the brightness evolution in \S \ref{sss:brightness}, strong conclusions
cannot yet be drawn.

%% file: issigi_larger.tex
\section{Collisions at larger separations}
\label{ss:larger}

While giant impacts are perhaps most associated with terrestrial planet formation,
there is substantial evidence from the Solar System that such events can take place
in the outer reaches of planetary systems too (\S~\ref{sss:outerss}).
Debris from giant impacts at larger separations undergoes the same evolution as
described in \S \ref{ss:gideb}.
The one point of note is that, for a given progenitor, $\sigma_v/v_{\rm k}$ will be
higher (and so the debris ends up in a broader annulus) for impacts at larger
separations (see Fig.~\ref{fig:sigvvk-orbdist}).
However, the considerations for the detectability of debris from large separation
impacts are slightly different than for impacts closer to the star
\citep[see][]{Jackson2014}.

One obvious difference is that dust is colder at larger separations,
and so it is much more difficult to detect its emission at 12\,$\mu$m.
This is compounded by the more general point that a larger mass of dust (and so a larger
progenitor mass) is required for a detectable level of emission at larger separations.
The emission from giant impact debris at 10s of au would be most readily
detected in the far-IR.
Cold dust is quite commonly detected around nearby stars, with $\sim 20$\% of
nearby Sun-like stars having far-IR emission from circumstellar debris
\citep[e.g.,][]{Eiroa2013}.
However, this is not interpreted as giant impact debris.
Rather, this dust is believed to be the product of steady state collisional grinding
of planetesimal belts that are analogous to the Kuiper Belt in the Solar System
\citep{Wyatt2007b,Lohne2008,Gaspar2013}.
At the large orbital separations inferred for these belts, collisional lifetimes
are long enough that their dust can persist several Gyr above the detection
threshold.
Since observations of how debris levels vary with stellar age are consistent with
the collisional erosion of Kuiper belt analogues, this interpretation is favoured
over a giant impact origin for the dust \citep{Wyatt2007b}.
Thus it is not possible to use the presence of cold dust from photometry to
infer a recent giant impact as it was for hot dust found closer to the star
\citep{Kenyon2005}.

However, as is evident in Fig.~\ref{fig:moondynsnaps}, the asymmetric
morphology of a debris disk may be used to point to a recent giant impact.
The detectability of such asymmetries is favoured at large separations, because
the orbital timescales there are much longer than those in the terrestrial planet 
region, which means that the early asymmetric phases of the disk evolution last
much longer.
For example, while in the terrestrial planet region the typical lifetime of the
asymmetric disk is a few thousand years, this lifetime may rise to of order a million
years at 10s of au.
In addition it is easier to resolve the morphology of debris in the outer reaches
of a system due to its larger angular size (notwithstanding the longer wavelength
of the emission).
The most intuitive evidence of a giant impact would be the detection of an expanding
clump \citep[e.g.,][]{Wyatt2002}.
However, the clump phase is so short-lived that it is much less likely to be witnessed
than the asymmetric disk phase which is much longer lived.

In addition to the asymmetry in the distribution of debris seen in Fig.~\ref{fig:moondynsnaps},
the dominance of the collision-point in setting the collision rate of the debris 
means it plays a further important role in enhancing the asymmetry observed during the
asymmetric disk phase at wavelengths that trace the distribution of certain collision products.
Such products include the very smallest dust grains that are placed on orbits that are
significantly modified by radiation pressure (i.e., those close to or below the blow-out limit
given by eq.~\ref{eq:dbl}).
This is because such grains are put on orbits with pericentres that are close to the site of
their production, which means that when such grains are closest to the star they are
predominantly found close to the collision-point (see Fig.~\ref{fig:collmap}).
This leads to an excess of hot emission at this point, which may stand out further
because such grains are also heated to temperatures above black body equilibrium (and so are
much hotter than the rest of the debris);
their apocentre distribution also means that their spatial distribution extends to much
larger distance on the side of the star opposite the collision-point.

Another collision product with an enhanced asymmetry due to the collision-point is carbon monoxide
(CO) gas.
Cold bodies at large distances from their host star may contain CO ice (e.g., Solar System
comets are typically $\sim$1-10\% CO by mass), and when they are broken up in collisions this
might be released as gas, promptly through vaporisation of the ice in the collision, or
perhaps delayed through desorption from freshly exposed surfaces.
CO gas has a comparatively short lifetime of around 150~years to photo-dissociation by
interstellar ultraviolet radiation \citep[e.g.][]{Visser2009}.
This is comparable to the orbital timescale at tens of au from the host star, and thus the CO
will undergo substantial decay over the course of the orbit after its production, which is
predominantly at the collision-point (see Fig.~\ref{fig:collmap}).
Thus the asymmetry in the CO distribution will be enhanced relative to the parent body
population, since not only does the disk geometry reduce the CO density away from the
collision-point, but so does the photo-dissociation.

\begin{figure}[!tb]
  \centering
  \includegraphics[width=0.8\textwidth]{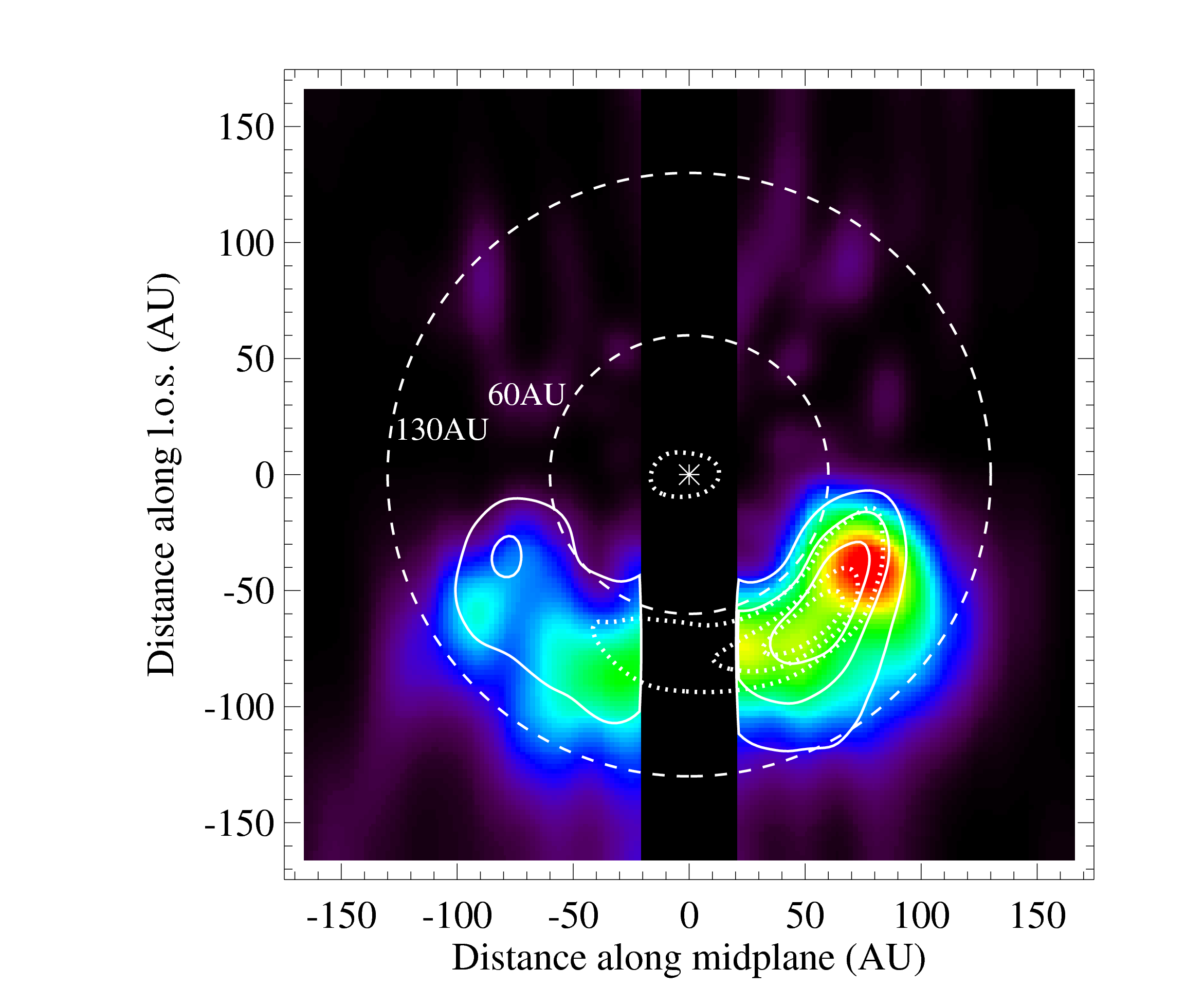}
  \caption{Image of the face-on distribution of CO gas in the $\beta$ Pic disk, derived from  
  spatial and velocity information in ALMA observations of this edge-on disk \citep{Dent2014}.
  Note this is just one possible de-projection, since the observations
  are insensitive to reflections in the $x$-axis (i.e., the gas has the
  same line-of-sight velocity whether it is in front or behind the star).
  The dashed circles denote the inner and outer radii of the gas disk at 60-130\,au,
  and the asterisk represents the location of the star.
  The distance along the line of sight (l.o.s.) is relative to the stellar position and is
  positive for gas on the far side of the star;
  orbital motion is clockwise.
  Overlaid on the colour map of the observed CO are contours of a model for CO produced at
  the collision-point of a giant impact involving Mars-size embryos at 85~au \citep{Jackson2014}.
  The solid contours show the model distribution after being run through the same
  de-projection process as the ALMA data, while the dotted contours show the true
  distribution.}
  \label{fig:betapicdecon}
\end{figure}

One system in which this effect might be visible is the well known edge-on disk of $\beta$ Pic.  
\citet{Telesco2005} revealed a large brightness asymmetry in the SW portion of the disk
at a projected separation of 52\,au in the mid-infrared, and \citet{Dent2014} showed that this
asymmetry is also present at a lower level in sub-millimetre continuum images.
\citet{Dent2014} also found that an asymmetry is present in CO line observations at the same
location, but with a significantly greater contrast.
The CO line observations also provide information on the line-of-sight
velocity of the gas, allowing the edge-on images of the gas distribution to be de-projected
to get a face-on view of the disk, under the assumption that the gas is in Keplerian
rotation.
Fig.~\ref{fig:betapicdecon} shows one possible de-projection in which most of the CO is in
a clump at around 85\,au from the star, with a tail extending 1/3 of an orbit in front
of the clump \citep{Dent2014}.
In optical scattered light the asymmetry is reversed, with the dust extending
around 25\% farther in the NE compared with the SW \citep{Larwood2001}.

While it is possible that the clump in $\beta$ Pic was created in the break-up of a planetesimal
within the last 50 years \citep{Telesco2005}, we are many orders of magnitude more
likely to witness the same impact in the asymmetric disk phase than in the clump phase.
Moreover the morphology, and in particular its wavelength dependence outlined above, are
exactly that expected if the asymmetry in the clump is the result of a giant impact witnessed
in the asymmetric disk phase;
e.g., Fig.~\ref{fig:betapicdecon} shows contours corresponding to a model for CO gas produced
in a giant impact onto a Mars-sized progenitor occurring at 85~au from the star
\citep[based on the model presented in][]{Jackson2014}.
However, an alternative explanation exists for the asymmetry in the
$\beta$ Pic disk that invokes a population of debris trapped in resonance with an unseen 
planet \citep{Dent2014}.
Such resonant populations have a clumpy distribution \citep[e.g.][]{Wyatt2003}, and the
collision rates would be enhanced in the clumps \citep{Wyatt2006}, leading to similar
asymmetries in the distributions of both CO and small grains to those of the asymmetric
disk phase of a giant impact.

One way to distinguish between the two interpretations is to look for motion of the clump relative
to the star, since the collision-point in the giant impact model is stationary on orbital
timescales (only moving on much longer precession timescales), whereas resonant clumps would move
with the planet.  
\citet{Li2012} re-imaged the mid-infrared clump and found tentative evidence for motion,
but further observations are needed to definitively confirm or reject its motion.
If the clump were found to be stationary, and so the giant impact interpretation favoured, this
would have significant implications for our understanding of planet formation processes in
the outer reaches of planetary systems.
This would imply that sufficient Mars-sized embryos exist at a time of $\sim 20$\,Myr after the
star formed for us to be likely to witness a collision between them in a $\sim 10$\,Myr timeframe.
This would set constraints on the total mass of such embryos and how they are distributed
in the disk and could be compared with models for the formation of such embryos
\citep[e.g.,][]{Kenyon2008}.
It would also boost the prospects for observing such collisions in other disks, as has been
already been proposed for HD181327 \citep{Stark2014}.

%% file: issigi_superearth.tex
\section{Giant collisions in super-Earth systems}
\label{ss:superearth}

As pointed out in \S~\ref{ss:larger}, while giant impacts are an integral part of the process 
of forming bodies like our own terrestrial planets, they are not exclusive to the formation of Solar 
System-like terrestrial planets.
One of the most interesting outcomes of exoplanet searches in recent years is the emergence of
a new class of {\it super-Earth} planets.
Such planets lie between roughly Earth and Neptune in size and/or mass and orbit in the inner
regions of their planetary systems.
There are no analogues to these planets in our Solar System, and yet they make up a substantial 
portion of the population of exoplanets discovered to date, thought to exist around
30-50\% of stars (Howard et al. 2010; Mayor et al. 2011).
As such the formation mechanism of these planets is the subject of
intense study and debate.

The key question regarding the origin of super-Earths is whether they formed in-situ
at $\ll 1$\,au \citep{Hansen2012,Hansen2013,Chiang2013}
or at larger distance then migrated in through interaction with the protoplanetary
disk \citep{Alibert2006,Lopez2012,Kenyon2014}.
Knowing the internal composition of the planets would help, e.g. with a more distant
formation location favoured for volatile-rich planets.
A significant question is thus whether these planets are `super-Earths'
that are largely rocky, or `sub-Neptunes' that are rich in volatile elements.
It is becoming clear that planets with radii larger than around 1.5~$R_{\oplus}$
cannot be purely rocky \citep[e.g.][]{Weiss2014, Rogers2015}, however the nature and
quantity of the lighter component of planets inferred to have low density is unclear.
Planets with radii of around 2-3~$R_{\oplus}$ can be well fit by a rocky core with a 
hydrogen-helium envelope that comprises $\sim$0.1-5 per cent of the planet mass 
\citep[e.g.][]{Wolfgang2015}, however they could equally well have less hydrogen and
helium, but be predominantly composed of water \citep[e.g.][]{Seager2007, Howe2014}.
Thus far atmospheric studies of such planets have been inconclusive
\citep[e.g.][]{Knutson2014, Kreidberg2014}.

This section explores the possibility that constraints can be set on the origin and
evolution of super-Earth planets by considering the debris released in giant impacts involving
such planets.
If rocky super-Earths form in situ in the same way as smaller terrestrial planets, they would
be expected to undergo giant impacts with concomitant debris releases during their formation.
A super-Earth planetary system formed by inward migration may also undergo a late phase of
giant impacts after the protoplanetary disk dispersed, for example if this promoted instability
in the system.
Regardless of its origin a super-Earth planetary system may also undergo late giant impacts
with any remnant planetary embryos in the system.
While the rate of giant impacts experienced by a super-Earth system has yet to be discussed
in the literature, the discussion in \S \ref{sss:common} shows that it should at least
possible to set constraints on that rate from observations.

To focus the discussion of giant impact debris in super-Earth systems, here we consider
the HD~69830 system.
This system contains three super-Earth planets with minimum masses of 10.2, 11.8 and
18.1~$M_{\oplus}$ at orbital distances of 0.079, 0.19 and 0.63~au respectively
\citep{Lovis2006} as well as a bright hot debris disk \citep{Beichman2005}.
Modelling of the dust emission spectrum suggests that the debris is centred at
around 1~au \citep{Lisse2007}, though this indirect constraint on its radial
location has a factor of a few uncertainty;
direct measurements of the radial location are consistent with a 1~au location,
but again have a factor of a few uncertainty \citep{Smith2009b}.
This system is particularly interesting as collisional modelling indicates that the dust
cannot be produced by the steady-state grinding of a planetesimal belt that has existed at 1\,au
since the formation of the system $\sim$2~Gyr ago, but rather must have been created more recently
\citep{Wyatt2007b,Heng2010}.
The fact that the predicted location of the debris disk is very close to, and could be coincident
with, the outermost planet HD~69830d also points to the possibility that the two are related.

\begin{figure}[!htb]
 \includegraphics[width=\textwidth]{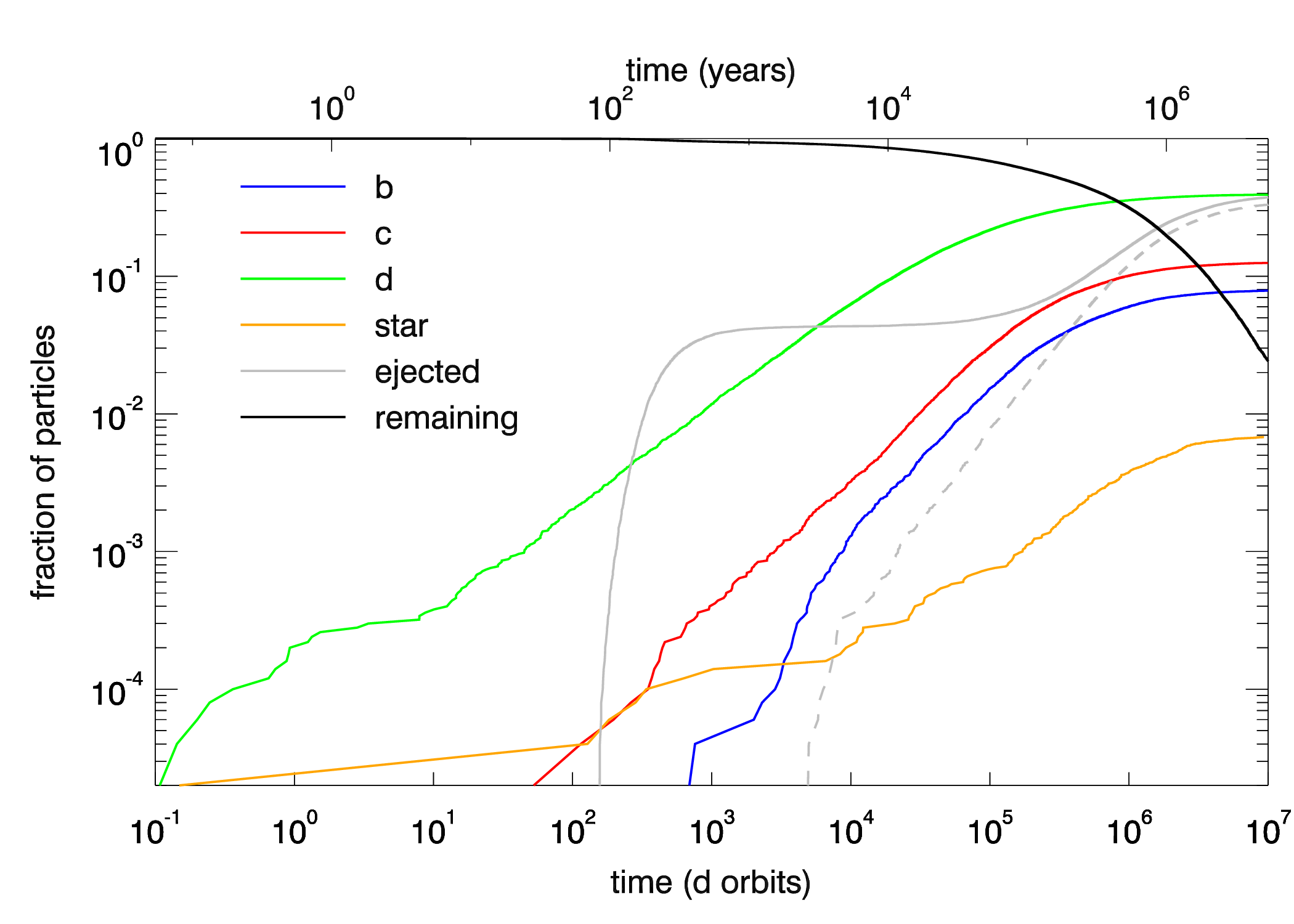}
 \caption{Fate of particles in a 50,000 particle $N$-body simulation of debris released in a giant 
  impact onto the planet HD~69830d.
  Time is given in terms of d orbital periods (197 days).  
  By 10$^7$ d orbits after the impact $>$95 per cent of the debris has been lost.
  Both dashed and solid grey lines correspond to particles that are ejected, however for
  the dashed line the prompt ejections (i.e., those particles placed on unbound orbits
  in the impact itself) are removed to make the longer term evolution more visible.
  The black line shows the number of particles remaining in the simulation.}
\label{fig:HD69830accretion}
\end{figure}

While there are other explanations for the dust that do not involve HD~69830d,
such as a recent collision within a stable asteroid belt at 1\,au \citep{Wyatt2010}
or a super-comet \citep{Beichman2005}, consideration of giant impact debris from
HD~69830d will highlight the differences and similarities to terrestrial planet
giant impact debris that will be useful for a more general discussion of the
detectability of super-Earth giant impact debris.
Fig.~\ref{fig:HD69830accretion} shows results from a simulation of the dynamical evolution
of debris released by a hypothetical impact involving HD~69830d.
For the purposes of determining the velocity dispersion of the debris it was assumed that
the planet has the same density as Neptune (1.64~g/cm$^3$), which leads to
$\sigma_v/v_{\rm k}=0.32$.
This shows that accretion onto all of the planets (b, c and d) is rapid, as is ejection,
with around 98\% of the particles having been lost by 10$^7$ d orbits (5\,Myr) after
the impact (with roughly twice as many lost to accretion as to ejection).
This contrasts with the Solar System case of debris from Moon-formation where only
44\% of the particles had been lost to dynamical effects at 10~Myr
(see Fig.~\ref{fig:moondebacc}).

A substantial part of this difference is due to the larger radii of the HD~69830 planets, 
since a Neptune-like HD~69830d has a radius of almost 4~$R_{\oplus}$ which would
result in an accretion rate 16 times larger than Earth (Eq.~\ref{eq:racc}, noting that
re-accretion onto progenitors has a similar gravitational focussing factor).
A more Earth-like composition, with a higher bulk density, would result in a smaller radius
and a lower accretion rate, and so a longer dynamical evolution than shown in
Fig.~\ref{fig:HD69830accretion}.
Another difference with debris from the Moon-forming impact is that
the accretion curves of all three planets around HD~69830 are roughly parallel.
This is because the larger mass (and escape velocity) of HD~69830d compared with
Earth leads to the debris being much more broadly distributed across the system.
As a result, and due to the relatively closely-packed nature of the system, all three planets
interact strongly with the debris from the beginning of the simulation.
A higher density for HD~69830d would enhance this effect due to the corresponding
increase in $\sigma_v$.

In addition to the more rapid dynamical evolution for super-Earth impact debris, the
collisional evolution will also be faster, since the launch (and so collision) velocities
will be higher than for debris from an Earth-like planet.
A larger fraction of the debris may also be released as vapour, since the larger escape
velocities of super-Earths imply that material ejected in their giant impacts will have
been subject to more violent shocks.
Both of these effects would add to the tendency for debris from a super-Earth giant impact
to last for less time than one involving an Earth-size planet.
While the initial mass of debris released from a more massive planet will be larger,
this cannot offset the more rapid collisional evolution of debris from a super-Earth giant impact
and so the duration of detectability would be unaffected, though this would mean that the disk
is brighter while it is present.

This section only scratches the surface of the topic of giant impacts in systems 
of super-Earths.
However, it is clear that there are potential implications for our 
understanding of the formation and composition of these worlds.
For example, eq.~\ref{eq:fgi} can be rewritten as
\begin{equation}
  t_{\rm gise12} = 100\,{\rm Myr} * f_{\rm 12} * f_{\rm 12gise} / f_{\rm se},
  \label{eq:tgise12}
\end{equation}
where $f_{\rm se} \approx 0.3$ is the fraction of stars that form super-Earths,
$f_{\rm 12} \approx 0.03$ is the fraction of 10-120\,Myr stars that have 12\,$\mu$m excess,
$f_{\rm 12gise} < 1$ is the fraction of those 12\,$\mu$m excesses that arise from
giant impacts onto super-Earths, and $t_{\rm gise12}$ is the duration of detectability
of giant impact debris in the 10-100\,Myr window from super-Earths at 12\,$\mu$m.
This calculation benefits from several parameters being well constrained,
leading to $t_{\rm gise12} <10$\,Myr.
This excludes a long period of late giant impacts which make long-lived detectable debris,
which could suggest most of the formation processes takes place in the
protoplanetary disk, and/or that only a small fraction of the impact-generated debris
is in the form of boulders.
However, some systems with infrared excess have already been interpreted as originating in
super-Earth giant impact debris \citep{Zuckerman2012}, which if confirmed could indicate
that such debris persists for a non-negligible duration (e.g., $t_{\rm gise12}>0.1$\,Myr).

%% file: issigi_conclusions.tex
\section{Conclusions}
\label{s:giconc}

There is abundant evidence for giant impacts in the history of the Solar System.  There are many ways in which such impacts 
could be manifested in observations of extrasolar systems, including in the properties of any extrasolar planets (e.g., their 
spin, surface, density, atmosphere, and moons), but the most promising method at the moment for detecting giant impacts is 
through debris released in the collisions and the resulting infrared emission from dust that is heated by the star.  This 
dust is likely to be readily detectable long after the event, up to 10-40\,Myr in the case of the Earth-Moon forming impact.  
However, there remain significant uncertainties in the duration of detectability because of the unknown size distribution of 
impact-generated debris.

There are 10s of candidate young Sun-like stars with bright mid-IR emission from dust at levels expected following a giant 
collision like that which created the Moon.  One of the best candidates for giant impact debris is around the star HD172555, 
since its spectrum shows dust with a silica composition that suggests an origin in a hypervelocity impact.  Other systems 
also show complex variability in their emission, similar to that expected due to the dynamical and physical evolution of 
impact generated debris, which could provide a valuable window into these events.  Despite the growing number of examples of 
potential giant impact debris, the occurrence rate of such emission is relatively low, at a few \%, which either implies that 
late stage giant impacts are rare (i.e., planet formation is largely complete by the time the protoplanetary disk has 
dispersed), or that the debris from impacts rapidly grinds itself down to undetectable levels.

Giant impacts in terrestrial planet forming regions are those most readily identified, since these inner regions are usually 
clear of dust following the dispersal of the protoplanetary disk.  However, giant impacts may also occur at larger distances 
from the stars (at 10s of au).  At such distances, larger quantities of debris are required to be detected photometrically, 
not least because emission from cold debris is relatively common at this distance ($\sim 20$\% for Sun-like stars), so 
evidence for giant impacts must rest on the morphology of the disk.  A clump observed in the $\beta$ Pictoris debris disk may 
be evidence of a collision between icy Mars-sized embryos.  If so this would require the outer regions of this system to be 
host to multiple such embryos to give a reasonable chance of witnessing a collision.  The ubiquitous class of super-Earth 
planetary system may also be subject to giant impacts. This chapter provides an initial exploration of this topic by showing 
how the dynamical evolution differs from that of Moon-forming debris, and it is shown that studying such impacts could 
provide important clues to the formation of super-Earth systems.